\documentclass[a4paper,11pt]{article}
\pdfoutput=1 

\usepackage{jcappub} 
                     
\usepackage{graphicx,amsfonts,amssymb,comment,amsmath,hyperref,float,times}
\usepackage{mathrsfs}  
\usepackage{amsmath,amssymb,dsfont}
\usepackage{cancel}
\usepackage[normalem]{ulem}
\usepackage{mathtools} 
\usepackage{multirow}
\usepackage[dvipsnames]{xcolor}
\definecolor{myBlue}{rgb}{0.24,0.34,0.58}
\definecolor{myRed}{rgb}{0.73,0.24,0.23}
\definecolor{myGreen}{rgb}{0.29,0.48,0.27}
\definecolor{myYellow}{rgb}{0.88,0.73,0.33}

\usepackage[font=small,labelfont=bf]{caption}

\usepackage{tikz}

\newcommand{\dd}{\; \mathrm{d}}
\newcommand{\isotope}[2]{${}^{#2}$#1}

\title{\textsc{DaMaSCUS}:\\ The Impact of Underground Scatterings on Direct Detection of Light Dark Matter}


\author{Timon Emken}
\author{and Chris Kouvaris}
\affiliation{$\text{CP}^3$-Origins, University of Southern Denmark, Campusvej 55, DK-5230 Odense, Denmark}
\emailAdd{emken@cp3.sdu.dk}
\emailAdd{kouvaris@cp3.sdu.dk}

\abstract{
Conventional dark matter direct detection experiments set stringent constraints on dark matter by looking for elastic scattering events between dark matter particles and nuclei in underground detectors. However these constraints weaken significantly in the sub-GeV mass region, simply because light dark matter does not have enough energy to trigger detectors regardless of the dark matter-nucleon scattering cross section.
Even if future experiments lower their energy thresholds, they will still be blind to parameter space where dark matter particles interact with nuclei strongly enough that they lose enough energy and become unable to cause a signal above the experimental threshold by the time they reach the underground detector.
Therefore in case dark matter is in the sub-GeV region and strongly interacting, possible underground scatterings of dark matter with terrestrial nuclei must be taken into account because they affect significantly the recoil spectra and event rates, regardless of whether the experiment probes DM via DM-nucleus or DM-electron interaction.
 To quantify this effect we present the publicly available Dark Matter Simulation Code for Underground Scatterings (\textsc{DaMaSCUS}), a Monte Carlo simulator of DM trajectories through the Earth taking underground scatterings into account. Our simulation allows the precise calculation of the density and velocity distribution of dark matter at any detector of given depth and location on Earth. The simulation can also provide the accurate recoil spectrum in underground detectors as well as the phase and amplitude of the diurnal modulation caused by this shadowing effect of the Earth, ultimately relating the modulations expected in different detectors, which is important to decisively conclude if a diurnal modulation is due to dark matter or an irrelevant background. \\[0.2cm]
\textit{Preprint: CP3-Origins-2017-20 DNRF90}
}

\begin{document}
\maketitle
\flushbottom

\section{Introduction}
\label{s:intro}
The direct detection of dark matter (DM) particles from the galactic halo has been an ongoing endeavour for the last three decades~\cite{Goodman1985,Drukier1986}. The various direct detection experiments have been the most straight forward strategy to shed light on one of the most intriguing questions in physics, the nature of dark matter. Yet no conclusive experimental evidence for DM has been found so far and we are still confronted with the discrepancy between the virtually conclusive gravitational evidence on all large scales, from galactic to cosmological~\cite{Bertone2004}, and the absence of any Earth-based experimental evidence.

The conventional direct detection approach is based on elastic DM-nucleus collisions and the subsequent observation of the nuclear recoil. Experiments such as LUX~\cite{Akerib2013,Akerib2014,Akerib2016} have been successful in constraining the standard `WIMP paradigm', putting severe bounds on interaction cross-sections for DM masses above several GeV. The continuing null results have therefore led to a shift towards the relaxation of underlying assumptions and a new focus on DM models beyond the classic WIMP.

One aspect of this shift is the redirection of experimental efforts towards lighter DM. Even though we do not make assumptions about UV-completions in this paper, there exists a series of models accommodating light DM, e.g. asymmetric DM~\cite{Nussinov1985,Kaplan1992,Gudnason2006,Gudnason2006a,Kaplan2009,Falkowski2011,Lin2012}. Sub-GeV weakly interacting particles evade the common direct searches due to their soft nuclear recoil energies falling below experimental recoil energy thresholds, typically of the order of keV. Experiments like DAMIC~\cite{Aguilar-Arevalo2016}, CRESST-II~\cite{Angloher2016}, EDELWEISS-III~\cite{Hehn2016} or CDMSlite~\cite{Agnese2016} pushed the limits of conventional detectors by using low-mass target nuclei and realizing recoil thresholds as low as $\mathcal{O}$(100eV), such that the sensitivity reaches to DM masses down to $m_{\chi}\approx 500$ MeV for CRESST-II. For even lower DM masses however new detection strategies are required, since the discrimination between soft nuclear recoils and background becomes a serious problem.

A very promising idea is the search for DM-electron scatterings, which lead to new detection signatures sensitive to masses below $\sim$GeV. Different approaches have been suggested, such as electron ionizations and excitations in atoms or semiconductors~\cite{Essig2012,Graham2012,Lee2015,Essig2016}. To trigger this kind of signals the kinetic energy of the DM particles needs to exceed binding energies of the order of only $\mathcal{O}$(10eV) for atoms and $\mathcal{O}$(eV) for semiconductors, rendering possible the discovery of light DM, provided that the detector is sensitive to such low energy deposits. Using this channel first limits on sub-GeV DM using DM-electron scatterings have been presented for XENON10~\cite{Essig2012a} and XENON100~\cite{Essig2017}. Other proposed targets for DM-electron scattering experiments were scintillators~\cite{Derenzo2016}, superconductors~\cite{Hochberg2016,Hochberg2016a}, two-dimensional targets~\cite{Hochberg2016a} and superfluid helium~\cite{Schutz2016,Knapen2016}, some of which potentially sensitive to light DM with masses as low as $\mathcal{O}$(keV).

Apart from utilizing DM-electron scatterings, further search strategies include the use of the Bremsstrahlung of the nuclear recoil in a conventional detector~\cite{Kouvaris2016a}, allowing to constrain MeV-scale DM e.g. with xenon detectors~\cite{Kouvaris2016a,McCabe2017}. Additional methods exploit the disintegration of chemical bonds~\cite{Essig2016a}, or other effects~\cite{Bunting2017,Budnik2017}.

In light of these new ideas concerning the direct searches for sub-GeV DM, there is a crucial aspect that should not be neglected. While it is true that elastic DM-nucleon scatterings of light DM are not directly observable below a certain DM mass, it also means that the corresponding cross-section is still allowed to be large and scatterings on nuclei inside the Earth may occur nevertheless~\cite{Lee2015}. A prime example in the context of DM-electron scattering experiments are models which include a dark photon mixing kinetically with the SM photon~\cite{Letters1984,Holdom1986}. In the heavy mediator limit the DM-electron and the DM-proton scattering cross-section are related via
\begin{align}
	\frac{\sigma_{\chi p}}{\sigma_{\chi e}} \simeq \left(\frac{\mu_{\chi p}}{\mu_{\chi e}}\right)^2\, ,\label{eq:eN}
\end{align}
where $\mu_{XY}$ refers to the reduced mass of two particle species $X$ and $Y$. For DM masses of the order $\mathcal{O}$(100MeV) this leads to a interaction strength hierarchy of about $(\mu_{\chi p}/\mu_{\chi e})^2\sim(m_{\chi}/m_{e})^2\sim\mathcal{O}(10^4-10^5)$. Hence DM-electron scattering cross-sections being tested by new experiments are accompanied in this model by much stronger, yet unobservable, DM-proton interactions.

In cases like this, sub-GeV DM particles scatter elastically on terrestrial nuclei while they travel through the Earth towards the detector, leading to deformed DM density and velocity distributions due to deflection and deceleration, which in turn has an impact on any direct detection experiment regardless of its specific detection channel or search strategy.

The typical signature of this deformation are diurnal modulations of the detection signal rate. The average distance a DM particle moves through the Earth's bulk mass to reach a detector varies as the Earth rotates. Therefore the pre-detection scattering probability changes periodically over a sidereal day, and with it the modification of the DM density and velocity distribution. This is especially true for detectors in the southern hemisphere, which are much more sensitive to this `Earth shadowing' effect. This effect has been quantified in early MC simulations in the context of the classic WIMP~\cite{Collar1992,collar1993,Hasenbalg1997}, and further studied in the context of hidden sector DM~\cite{Foot2004,Foot2012,Foot2015} and DM-electron scattering experiments~\cite{Lee2015}. In the most extreme case underground detectors might lose detection sensitivity altogether, since the rock of the Earth crust above the laboratory, meant to serve as a background shield, effectively blinds the experiment via the nuclear stopping and screening of the incoming DM particles. This possibility has been studied in~\cite{Sigurdson2004,Zaharijas2005,Kouvaris2014} and more recently in the context of DM-electron scatterings~\cite{Emken2017}. Experimental efforts to observe a possible diurnal modulation have been carried out in the early 90s by the COSME-II detector~\cite{Collar1992} and later by the DAMA collaboration~\cite{Bernabei2014,Bernabei2015}. Both experiments were located in the northern hemisphere and failed to find any evidence for diurnal modulations. A promising future experiment in the southern hemisphere is the SABRE experiment at the Stawell Underground Physics laboratory~\cite{Shields2015,Froborg2016}, which is designed to test the DAMA/LIBRA discovery claim~\cite{Bernabei2008} and whose location would be significantly more sensitive to diurnal modulation due to Earth scatterings.

The detection signature of Earth scatterings has recently been investigated using analytic methods~\cite{Kavanagh2016}. Therein the authors published the \textsc{EarthShadow} code, which allows to quantify the lab-frame DM distribution distortions in the case, where the DM particles scatter at most once before passing through the detector. Therefore this approach is restricted to the single-scattering regime. For the case of multiple scatterings numerical methods become necessary. In this paper we present the Dark Matter Simulation Code for Underground Scatterings (\textsc{DaMaSCUS}), a Monte Carlo simulation code for individual DM trajectories, which allows to calculate local distortions of density and velocity distributions for any number of scatterings. In this sense \textsc{DaMaSCUS} complements and generalizes the \textsc{EarthShadow} code, which we use as a crucial consistency check for the new MC simulations. The simulations take the Earth's orientation in the galactic frame into account, as well as its composition and layer structure and time-dependent velocity through the halo, while it orbits the Sun. By simulating billions of particles we investigate how underground DM-nucleus scatterings affect the local DM velocity distributions by statistical means. This information may then be used to compute time dependent distortions of recoil spectra and diurnal modulations of event rates for any experiment while precisely accounting for its location, underground depth and search strategy. It is even possible, although computationally more expensive, to investigate the Earth's crust screening effect for very strong DM-nucleus interactions.

We should stress that this is an important effect that can play an important role if DM is light and sufficiently strongly interacting. Underground scatterings could render light DM completely unable to produce detectable nuclear recoils at the usual $\sim$ 1 km depths of most current detectors, leaving a hole in the DM parameter space which will not be covered neither with larger exposure nor with lower energy thresholds. Shallow-site or surface detectors looking for a diurnal signal might be the only working strategy for discovering DM in this part of the parameter space~\cite{Kouvaris2014}. 

This paper is organized as follows. After a brief review of basic DM-nucleus scatterings in section~\ref{s:basics} we describe the MC simulation algorithm of \textsc{DaMaSCUS} in section~\ref{s:MC}. Our main results are presented in section~\ref{s:results}, before we conclude and give an outlook on future steps in section~\ref{s:conclusion}. In addition we provide a set of extensive appendices for the interested reader, which contain short reviews of necessary astronomical relations and computational details of our simulations.

Together with this paper we also make the \textsc{DaMaSCUS} v1.0 code publicly available, together with documentation and some illustrative videos~\cite{Emken2017a}.

\section{DM Scatterings on Terrestrial Nuclei}
\label{s:basics}
The central physical process of interest are elastic collisions of DM particles on nuclei of the Earth's bulk mass. As the particles pass through the Earth's mantle and core they may interact with matter depending on their interaction cross-section. In this section we review the underlying dynamics and probabilities. First of all, in order to describe the DM particle's underground motion we need to model the Earth and its layer structure. The Earth's density increases with the depth, such that we have more nuclei near the core for the DM particle to scatter on. We implement the mass density profile of the Preliminary Reference Earth Model (PREM)~\cite{Dziewonski1981}, which separates the Earth into 10 distinct layers. Furthermore the chemical composition changes depending on the underground depth as well. We distinguish two different compositional layers, the core and the mantle, for each of which we implement the 9 and 14 most abundant nucleus species respectively~\cite{McDonough2013}. The details are summarized in appendix~\ref{a:earth}.

We consider a DM particle moving through matter. The probability for a particle of velocity $\vec{v}$ to scatter on some nucleus, after freely travelling a distance $L$, is
\begin{align}
	P(L)&=1-\exp \left[-\int\frac{\dd x}{\lambda_{\text{MFP}}(\vec{x},\vec{v})}\right]=1-\exp \left[-\int\limits_{0}^{L/v}\frac{\dd t}{\lambda_{\text{MFP}}(\vec{x}(t),\vec{v})}\frac{\dd x}{\dd t}\right]\, . \label{eq:scatterprobability}
\end{align}
\begin{figure}[tbp]
\centering
	\includegraphics[width=0.65\textwidth]{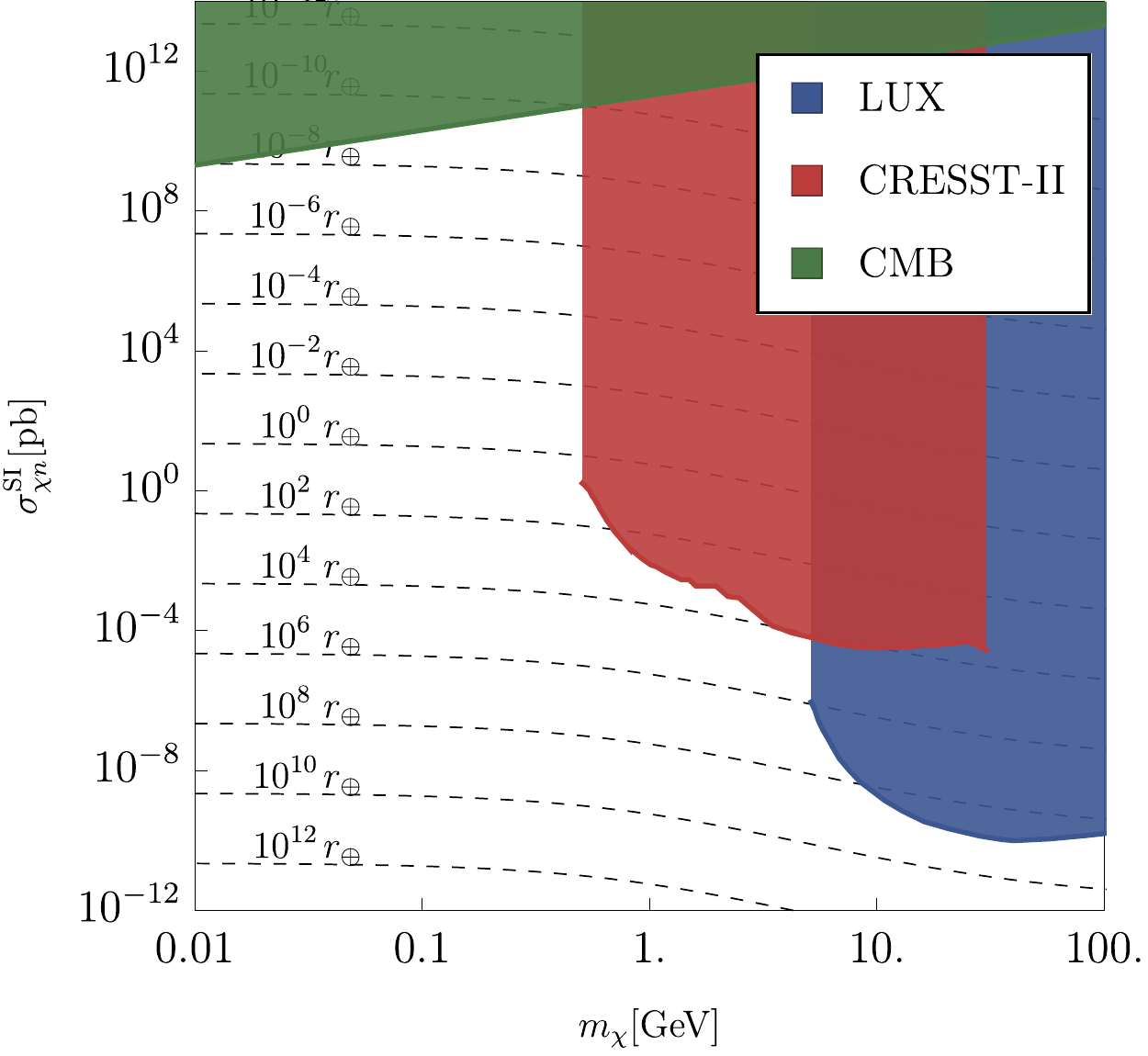}
	\caption{The dashed lines show the average underground mean-free-path $\overline{\lambda}_{\rm MFP}$ as a function of the DM mass and the spin-independent DM-nucleon scattering cross-section in units of the Earth radius. The direct detection constraints from LUX (WS2013+WS2014-16)~\cite{Akerib2016} and CRESST-II (2015)~\cite{Angloher2016}, as well as the constraints from the Cosmic Microwave Background~\cite{Dvorkin2013} show that for sub-GeV DM multiple underground scatterings on nuclei are still a very viable possibility.}
	\label{fig:mfp}
\end{figure}
In general the mean free path $\lambda_{\rm MFP}$ is a local and speed dependent property and given by
\begin{align}
\lambda_{\text{MFP}}^{-1} (\vec{x},\vec{v}) &=\sum_i \lambda_i^{-1} (\vec{x},\vec{v})\equiv \sum_{i}n_{A_i}(\vec{x})\sigma^{\rm total}_{\chi A_i}(\vec{v})\, ,
\end{align}
where we sum over all nuclear isotopes present at $\vec{x}$. $n_{A_i}(\vec{x})$ is the number density of the nucleus with atomic mass number $A_i$. Substituting the Earth's local chemical abundances, or the mass fractions of the different atomic species $f_{A_i}(\vec{x})$, as well as the density profile of the Earth $\rho_{\oplus}(\vec{x})$, we can rewrite Eq.~(\ref{eq:lambdainverse})
\begin{align}
\lambda_{\text{MFP}}^{-1} (\vec{x},\vec{v})&=\sum_{i}f_{A_i}(\vec{x})\frac{\rho_{\oplus}(\vec{x})}{m_{A_i}}\sigma^{\rm total}_{\chi A_i}(\vec{v})\, .\label{eq:lambdainverse}
\end{align}
In figure~\ref{fig:mfp} we show the average underground mean-free-path together with experimental constraints, which show that for sub-GeV DM the possibility of multiple Earth scatterings is still very viable. 

Suppose the DM particle scatters at $\vec{x}$, the probability to scatter on a certain nucleus species $j$ is given by
\begin{align}
	P(\text{scattering on $j$}) &=\frac{\lambda_j^{-1}(\vec{x},\vec{v})}{\lambda_{\rm MFP}^{-1}(\vec{x},\vec{v})}\, . \label{eq:scatternucleus}
\end{align}
 The scattering cross-section $\sigma_{\chi A}$ between a DM particle and a nucleus of mass number $A$ generally depends on the momentum transfer $q$ at least through a form factor that accounts for loss of coherence. For high momentum transfers with de Broglie wavelengths below the nucleus's size, DM cannot scatter coherently with all the nucleons composing the nucleus~\cite{Lewin1996}. This loss of coherence is taken care of by a form factor $F_A(q^2)$ via
\begin{align}
	\sigma_{\chi A}(q^2) = \sigma_{\chi A}(0)\; F^2_{A}(q^2)\, ,\quad \text{where } F_A(0)=1\, .
\end{align}
The total scattering cross-section is obtained, by averaging over all possible momentum transfers/recoil energies,
\begin{align}
	\sigma^{\rm total}_{\chi A} &=  \frac{\sigma_{\chi A}(0)}{q_{\rm max}^2}\int\limits_{0}^{q_{\rm max}^2}\dd q^2\; F^2_A(q^2)\, ,\quad\text{where }q_{\rm max}^2=4\mu_{\chi A}^2v_{\chi}^2\, . \label{eq:totalcs}
\end{align}
For sub-GeV DM it will not be necessary to take the loss of coherence into account\footnote{However an approximative Helm form factor is implemented in \textsc{DaMaSCUS} and may be used for the simulation of heavier DM.} since $F_A(q^2)\approx 1$, and hence $\sigma^{\rm total}_{\chi A} \approx \sigma_{\chi A}(0)$. But it is interesting to note that the form factor makes even the spin-independent total scattering cross-section velocity dependent via $q_{\rm max}$.

In modelling the DM-matter interactions we take the bottom-up framework of non-relativistic effective theory~\cite{Fitzpatrick2012a}. However in this work we only present results for the first operator, better known as spin-independent DM-nucleon interactions. A MC exploration of the other operators in analogy with~\cite{Kavanagh2016} will follow in a later publication. Hence for now we exclusively consider isospin non-violating spin-independent interactions, for which the zero momentum transfer cross-section is given as
\begin{align}
\sigma^{\text{SI}}_{\chi A}(0)=\sigma^{\text{SI}}_{\chi n}(0) \frac{\mu_{\chi A}^2}{\mu_{\chi n}^2}A^2\, ,\label{eq:sigmaN}
\end{align}
with the DM-nucleon cross-section $\sigma^{\text{SI}}_{\chi n}$ and the corresponding reduced mass $\mu_{\chi n}$.

\section{Monte-Carlo Simulations with \textsc{DaMaSCUS}}
\label{s:MC}
Having covered the basics we introduce the Dark Matter Simulation Code for Underground Scatterings (\textsc{DaMaSCUS}). \textsc{DaMaSCUS} performs simulations of individual particles traversing through the Earth's mantle and core undergoing scatterings on terrestrial nuclei, which deflect and decelerate the particle. It accounts for the changing composition and density throughout the Earth as well as its motion in the DM halo and its orientation in the galactic frame. A statistical sample of trajectories can be analysed to give precise estimates of the modified local DM number density and velocity distribution, which for any given direct detection experiment will be time-dependent. This allows to compute the local signal rate and its diurnal modulation for any specific experiment, similarly to the recent \textsc{EarthShadow} code~\cite{Kavanagh2016}. There diurnal modulations have been computed under the assumption that the DM particle scatters at most once on terrestrial nuclei before reaching a detector. Using MC simulations we are not restricted to the single scattering regime and can generalize these findings by simulating particles with any number of underground scatterings. We compare our MC simulation results with the ones of the \textsc{EarthShadow} in the single-scattering regime where the latter is valid as an extra consistency check of \textsc{DaMaSCUS}.

The basic idea of the MC simulation is to follow individual DM particle on their journey through Earth's interior, as they scatter on terrestrial nuclei resulting in trajectories not unlike a random walk. By recording how the scatterings diffuse DM particles underground, we can derive precise estimates of the DM density and velocity distributions at the location of detectors of interest.

As we will demonstrate below, the particles are sent underground with appropriate initial conditions $(t_{\rm ini},\vec{x}_{\rm ini},\vec{v}_{\rm ini})$ and assumed to move on a straight line until hitting a nucleus. In order to find the distance $L$ a particular particle travels freely we employ~\eqref{eq:scatterprobability} and solve the equation
\begin{align}
	P(L) = \xi \in (0,1)\, ,\label{eq:pxi}
\end{align}
where $\xi$ is a uniformly distributed random number. Then we define the displacement vector $\vec{\Delta}(\vec{x},\vec{v})$ as
\begin{align}
	\vec{\Delta}(\vec{x},\vec{v})=L\vec{e}_v\, ,
\end{align}
where $\vec{e}_v$ is the unit vector in the direction of $\vec{v}$. This vector points from the particle's original position $\vec{x}$ to the nucleus on which it scatters. The solution of~\eqref{eq:pxi} is found by an algorithm which combines analytic and numerical methods, for details we refer to appendix~\ref{a:displacement}.

Now that the location of the first scattering event is known, the particular nucleus $A$ involved in the scattering can be inferred from~\eqref{eq:scatternucleus}. The particle will deflect and decelerate and its resulting velocity after the scattering is given by a simple relation for elastic collisions,
\begin{align}
\vec{v}' = \frac{m_A \left|\vec{v}\right| \vec{n}+m_{\chi}\vec{v}}{m_A+m_{\chi}}\, .\label{eq:vnew}
\end{align}
Here the only unknown part is $\vec{n}$, the unit vector pointing into the direction of the DM particle's velocity after the scattering in the center-of-mass-frame. We define the scattering angle $\alpha = \sphericalangle(\vec{v}_{\chi},\vec{n})\in [0,\pi]$ as the angle between the incoming and outgoing direction of the DM particle in the CMS-frame. For spin-independent cross-sections there is no preferred value for $\alpha$. However we also note that by including the form factor, the scattering angle $\alpha$ will no longer be uniformly distributed even for the SI case. Instead we solve
\begin{align}
	\frac{\int\limits_{0}^{q^2}\dd q^2\; F^2_A(q^2)}{\int\limits_{0}^{q_{\rm max}^2}\dd q^2\; F^2_A(q^2)} = \xi \in (0,1)
\end{align}
for $q$, where $\xi$ again is a uniformly distributed random number. The scattering angle is then given by
\begin{align}
	\cos \alpha = 1-2\frac{q^2}{q_{\rm max}^{2}}\, .
\end{align}
In the case of sub-GeV DM, $q^2 \approx \xi q_{\rm max}^2$, and we obtain a uniform distribution in $\cos\alpha$.

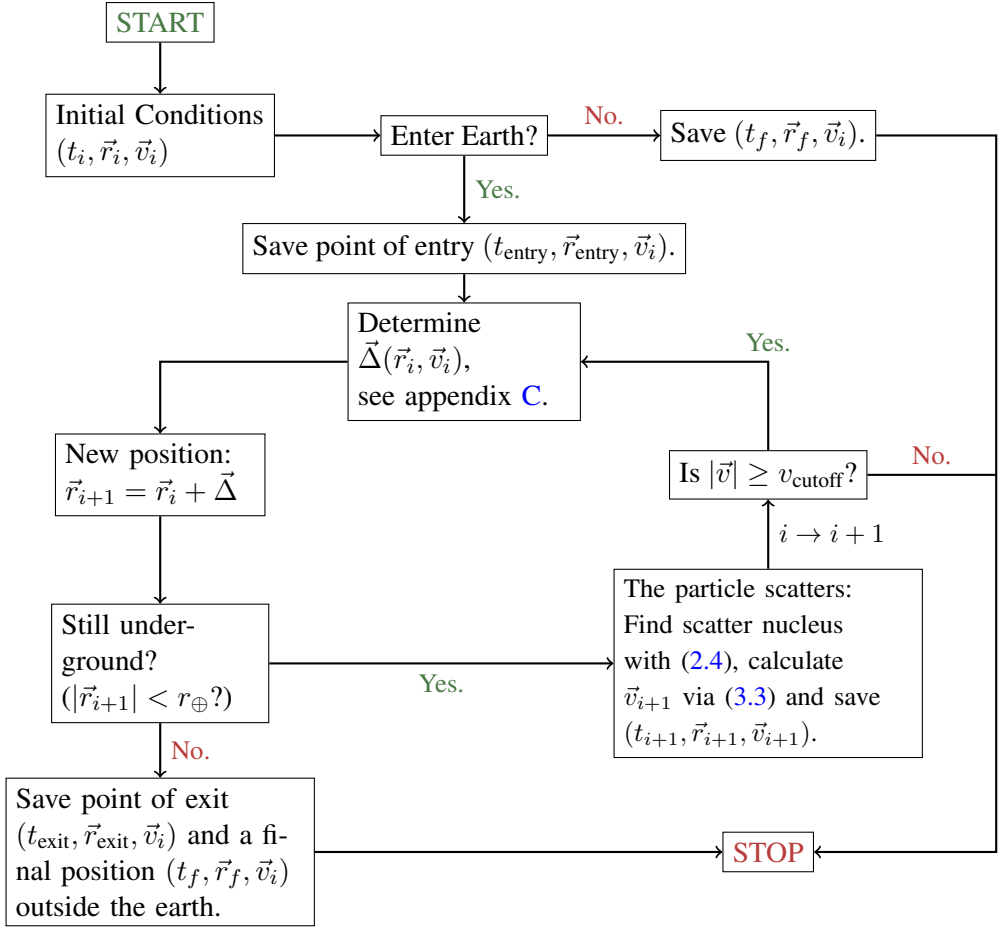
\begin{figure}[tbp]
\centering
\begin{tikzpicture}[scale=1.0]
\node[draw,rectangle] (a0) at (-8.0,0.0) {\color{myGreen}START};
\node[draw,rectangle,text width = 2.75cm](b0) at (-8.0,-1.5) {Initial Conditions $(t_i,\vec{r}_i,\vec{v}_i)$};
\draw[thick,->](a0.south)--(b0.north);
\node[draw,rectangle](b1) at (-4.0,-1.5) {Enter Earth?};
\draw[thick,->](b0.east)--(b1.west);
\node[draw,rectangle](b2) at (0.0,-1.5) {Save $(t_f,\vec{r}_f,\vec{v}_i)$.};
\draw[thick](b2.east)--(3.0,-1.5)--(3.0,-6.0);
\draw[thick,->](b1.east)--(b2.west)node[pos=0.5,above] {\small \color{myRed}No.};
\node[draw,rectangle] (c1) at (-4.0,-3.0){Save point of entry $(t_{\text{entry}},\vec{r}_{\text{entry}},\vec{v}_{i})$.};
\draw[thick,->] (b1.south)--(c1.north) node[pos=0.5,right]{\color{myGreen}\small Yes.};
\node[draw,rectangle,text width=2.8cm] (d1) at (-4.0,-4.5){Determine $\vec{\Delta}(\vec{r}_i,\vec{v}_i)$,\\ see appendix~\ref{a:displacement}.};
\draw[thick,->] (c1.south)--(d1.north);
\node[draw,rectangle,text width=2.5cm] (e2) at (-8.0,-6) {New position:\\$\vec{r}_{i+1}=\vec{r}_{i}+\vec{\Delta}$};
\draw[thick,->] (d1.west)--(-8.0,-4.5)--(e2.north) node[pos=0,above]{\small };
\node[draw,rectangle,text width=2.6cm] (f2) at (-8.0,-8.5){Still underground?\\ ($|\vec{r}_{i+1}|<r_{\oplus}$?)};
\draw[thick,->] (e2.south)--(f2.north);
\node[draw,rectangle,text width=3.8cm] (g2) at (-8.0,-11.0) {Save point of exit $(t_{\text{exit}},\vec{r}_{\text{exit}},\vec{v}_i)$ and a final position $(t_f,\vec{r}_f,\vec{v}_i)$ outside the earth.};
\draw[thick,->] (f2.south)--(g2.north) node[pos=0.5,right]{\small\color{myRed}No.};
\node[draw,rectangle](h2) at (0.0,-11.0) {\color{myRed}STOP};
\draw[thick,->](g2.east)--(h2.west);
\node[draw,rectangle,text width=3.8cm] (f1) at (0.0,-8.5) {\small The particle scatters:\\Find scatter nucleus with~\eqref{eq:scatternucleus}, calculate $\vec{v}_{i+1}$ via~\eqref{eq:vnew} and save $(t_{i+1},\vec{r}_{i+1},\vec{v}_{i+1})$.};
\draw[thick,->] (f2.east)--(f1.west) node[pos=0.5,below]{\small\color{myGreen} Yes.};
\node[draw,rectangle] (e1) at (0.0,-6.0) {Is $|\vec{v}|\ge v_{\text{cutoff}}$?};
\draw[thick,->] (f1.north)--(e1.south)node[pos=0.5,right]{\small $i\rightarrow i+1$};
\draw[thick,->] (e1.north)--(0.0,-4.5)--(d1.east) node[pos=0,above]{\small\color{myGreen} Yes.};
\draw[thick] (e1.east)--(3.0,-6.0)node[pos=0.5,above]{\small\color{myRed} No.};
\draw[thick,->] (3.0,-6.0)--(3.0,-11.0)--(h2.east);
\end{tikzpicture}
\caption{Flow chart for the Monte Carlo simulation algorithm of a single DM trajectory. The velocity cut-off $v_{\rm cutoff}$ is chosen very low ($\sim$ cm/s) and introduced to avoid numerical problems and save computation time for simulations with high DM-nucleon scattering cross-section.}
\label{fig:algorithm}
\end{figure}

The procedure repeats itself as the DM particle continues with its new velocity from the position of the scattering. Again we have to solve~\eqref{eq:pxi} and find the next scattering location and nucleus, and so on. This is repeated until the particle reaches the Earth surface again or until the velocity of the particle drops below a given speed threshold. The algorithm is summarized by a flow chart in figure~\ref{fig:algorithm}. It returns a list of events, which make up the DM particle's trajectory.

The choice of initial conditions for the simulated DM particles is critical. To find the initial time is trivial, we can set $t_{\text{ini}}$ to a random value or just start at $t_{\text{ini}}=0$. The initial velocity is straight forward as well and has two components,
\begin{align}
	\vec{v}_{\text{ini}} = \vec{v}_{\text{halo}}-\vec{v}_{\oplus}(t)\, .\label{eq:vini}
\end{align}
The first term is the velocity component in the galactic rest frame $\vec{v}_{\text{halo}}$, for which we choose the Standard Halo Model (SHM),
\begin{align}
	f_{\text{halo}}(\vec{v}) &= \frac{1}{N_{\text{esc}}} \exp\left(-\frac{\vec{v}^2}{v_{0}^2}\right)\Theta(v_{\text{esc}}-|\vec{v}|)\, ,\label{eq:fhalo}
\end{align}
where $N_{\text{esc}}=\pi v_{0}^2 \left( \sqrt{\pi}v_{0} \text{Erf}\left( \frac{v_{\text{esc}}}{v_{0}}\right)-2v_{\text{esc}}\exp \left(-\frac{v_{\text{esc}}^2}{v_{0}^2} \right)\right)$ is the normalization constant and $\Theta(x)$ the Heaviside step function. We use the standard parameter $v_0=220$km/s and $v_{\rm esc}=544$km/sec. The second component is the Earth's velocity $\vec{v}_{\oplus}(t)$ relative to the galactic rest frame, causing the ``DM wind''. This velocity is time dependent and changes over the course of one year, giving rise to annual modulation of detection signals~\cite{Drukier1986,Freese1988}. For details on its determination we refer to appendix~\ref{a:astro}.
\begin{figure}[tbp]
\centering
	\includegraphics[width=.45\textwidth]{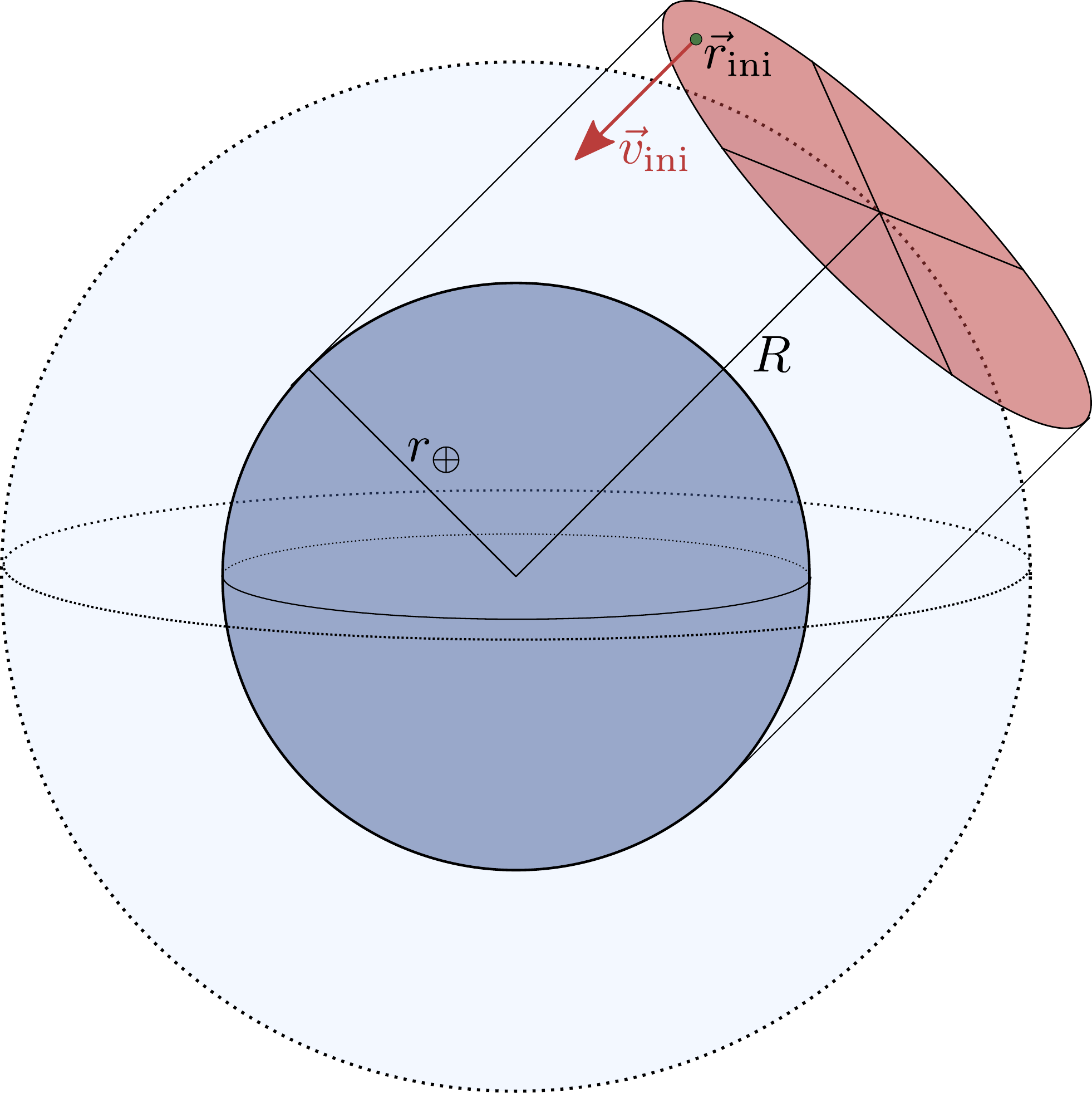}
	\caption{Sketch for the choice of initial position of the simulated DM particles. In order to secure an effectively uniform distribution in space, the particles are sent off from a random position on a disc of radius $r_{\oplus}$.}
	\label{fig:initialpositions}
\end{figure} 

Once the initial time and velocity are determined, the question of the initial position is more delicate. For one of course, the particle should start its trajectory outside the Earth and move towards the planet. But more importantly, the initial positions have to be distributed uniformly in space. They can not be chosen arbitrarily, e.g. simply on top the Earth's surface as done in~\cite{Collar1992,collarthesis1992,collar1993,hasenbalgthesis1994,Hasenbalg1997}. This point is subtle but crucial. Simulating particles with initial velocities given by~\eqref{eq:vini}, which exclusively start their trajectory on the surface of the Earth, creates a finite volume bias at shallow depths, i.e. exactly where detectors are located. Over proportionally many particles are sent into the Earth with narrow angles. We have checked that in the case of a transparent Earth, the aforementioned choice of initial conditions results in a DM over density close to the surface, in contrast to the expected uniform DM density. Apart from the numerical verification, it can also be shown analytically that the aforementioned choice of initial conditions does not describe the real situation.

Instead an effectively uniform distribution of the initial positions is realized by choosing a random point on a circular disk of radius $r_{\oplus}$ at a distance $R$ from the Earth center and perpendicular to $\vec{v}_{\text{ini}}$, see figure~\ref{fig:initialpositions}.
\begin{align}
	\vec{r}_{\text{ini}}&=R \vec{e}_z+\sqrt{\xi}r_{\oplus}\left( \cos \phi\; \vec{e}_x+\sin \phi \;\vec{e}_y\right)\, ,
\end{align}
where $\xi\in(0,1)$ and $\phi\in[0,2\pi)$ are uniformly distributed random numbers, and $\vec{e}_x$ and $\vec{e}_y$ span the disc. Together with a random starting time $t_{\text{ini}}$, this is equivalent to the choice of a random point inside a cylinder with radius $r_{\oplus}$, orientated parallel to $\vec{v}_{\text{ini}}$. This ensures the effectively uniform distribution of initial positions in space.

\begin{figure}[tbp]
	\centering
	\begin{minipage}[b]{0.5\textwidth}
		\includegraphics[width=0.95\textwidth]{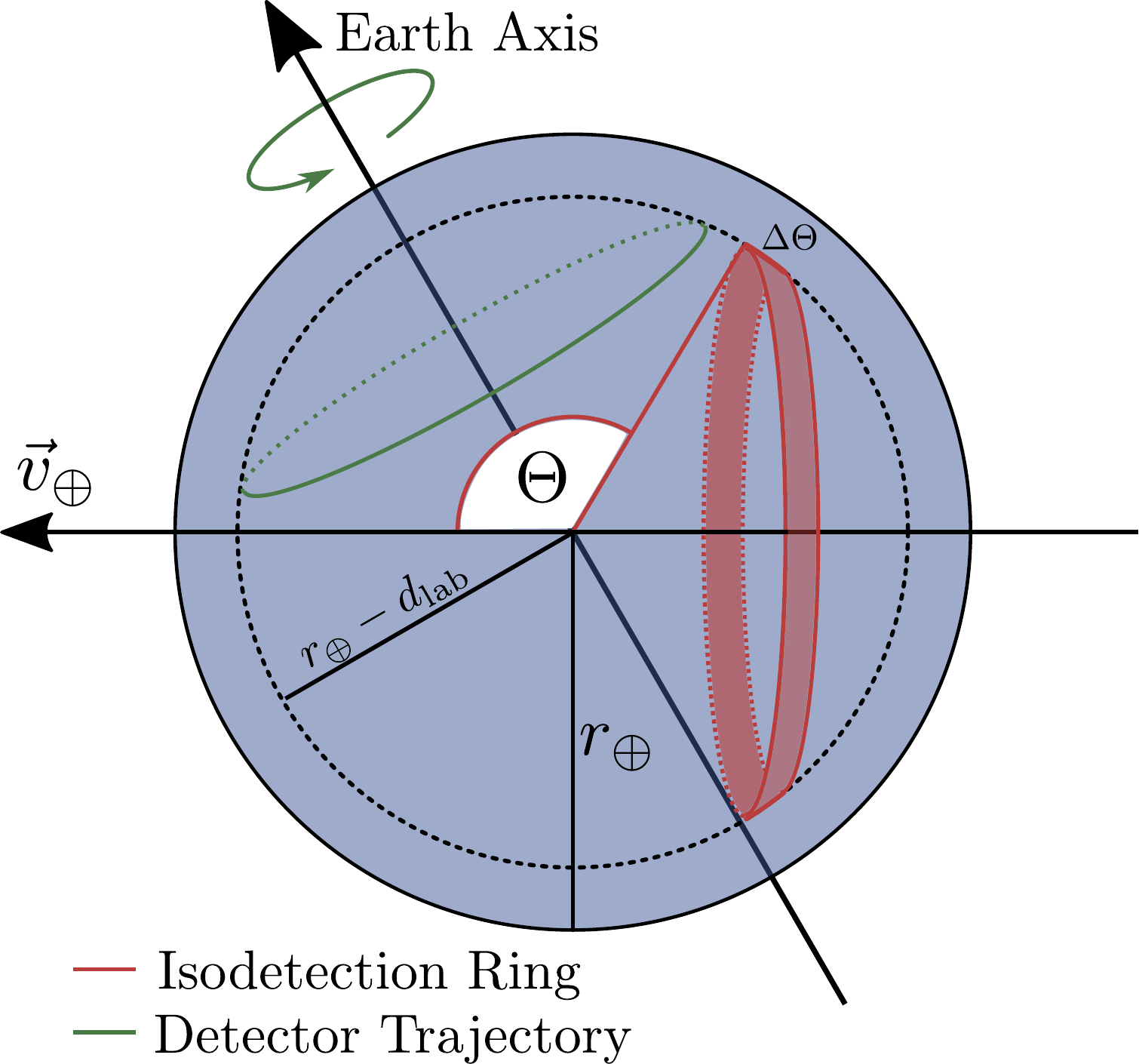}
	\end{minipage}
	\begin{minipage}[b]{0.45\textwidth}
		\includegraphics[width=0.95\textwidth]{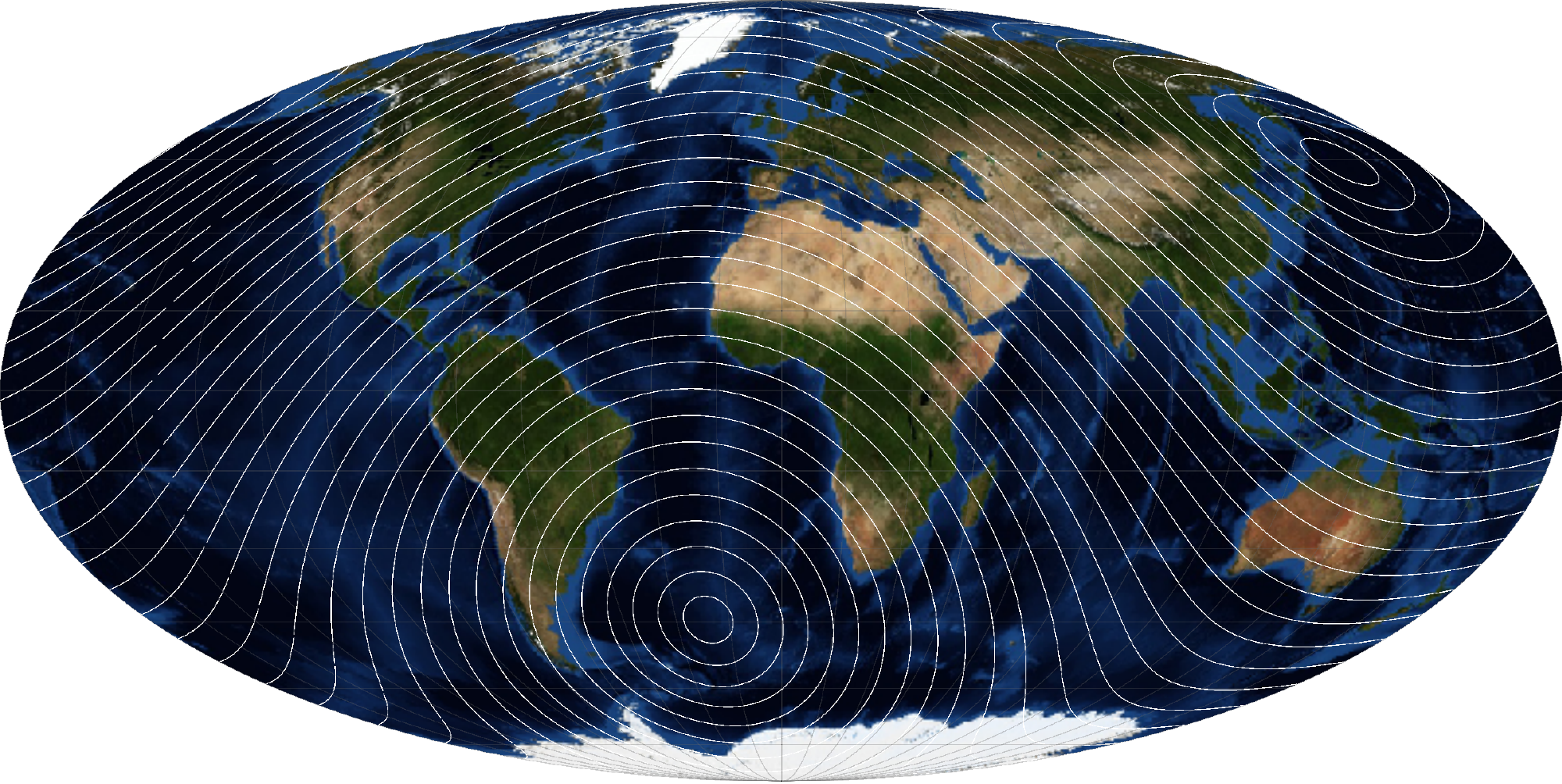}
		\includegraphics[width=0.95\textwidth]{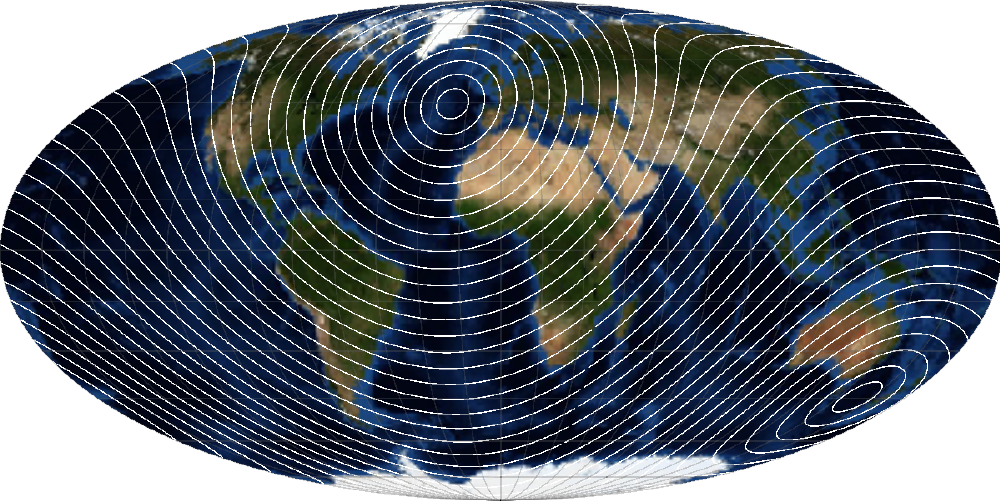}
	\end{minipage}
	\caption{Isodetection rings and their projection onto the Earth surface at 0:00 and 12:00. Here we chose $\Delta\Theta = 5^{\circ}$ for illustrative purposes.}
	\label{abb:isoringsmollweide}
\end{figure}
The halo DM velocity distribution~\eqref{eq:fhalo} is isotropic. Boosting our system into the frame of the Earth via~\eqref{eq:vini} breaks the isotropy, but nonetheless our system still has a rotational symmetry around the axis parallel to the Earth's velocity $\vec{v}_{\oplus}(t)$. We call the polar angle of this axis isodetection angle $\Theta$. Note that in~\cite{Kavanagh2016} and in the \textsc{EarthShadow} code the authors use an equivalent angle defined as $\gamma=180^{\circ}-\Theta$. As the name suggests, along a constant isodetection angle the DM particle's velocity distribution as well as direct detection event rates will also be constant. We exploit this symmetry for our MC simulations to define small but finite-sized isodetection rings as done in~\cite{Collar1992,collar1993,Hasenbalg1997}, see figure~\ref{abb:isoringsmollweide}. However we place these rings not at the Earth's surface, but at a finite underground depth, which can be adjusted for different experiments e.g. to 1400 meter for the LNGS. This way the MC simulation can also be used to investigate the Earth's crust shielding effect for very large interaction cross-sections. The effect of nuclear stopping on the sensitivity of underground direct detection experiments has been studied in~\cite{Zaharijas2005,Kouvaris2014,Kouvaris2016} and lately in the context of DM-electron scattering experiments using MC simulations in~\cite{Emken2017}.

\begin{figure}[tbp]
	\centering
	\includegraphics[width=0.7\textwidth]{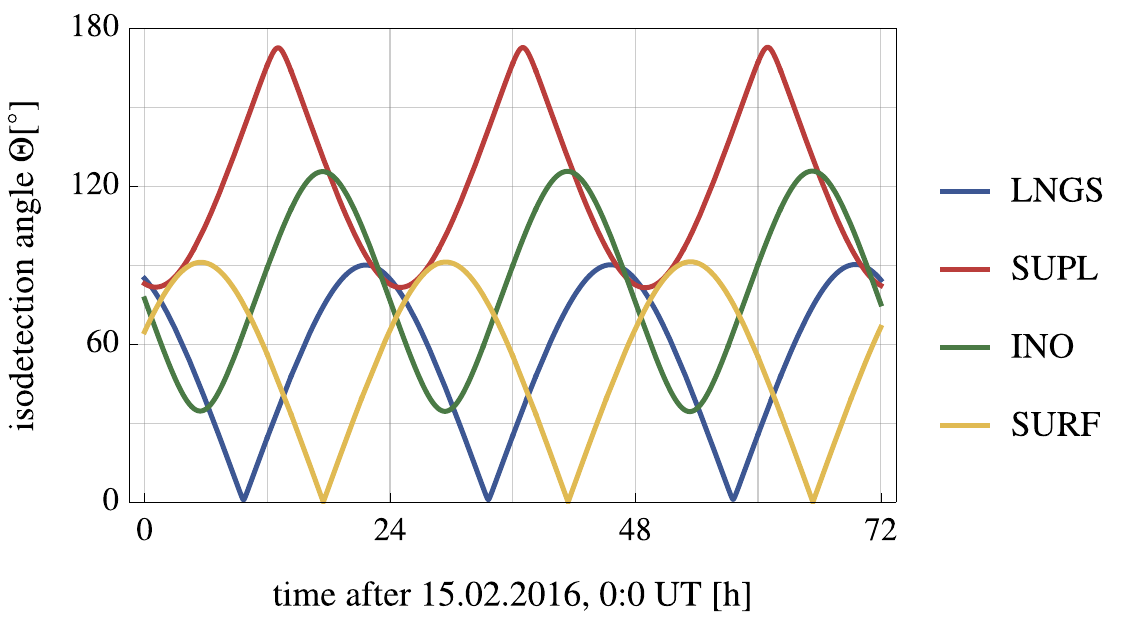}
	\caption{Isodetection angle for different laboratories over the duration of three days: the LNGS $(45.454^{\circ}N,13.576^{\circ}E)$, SUPL $(37.07^{\circ}S,142.81^{\circ}E)$, INO $(9.967^{\circ}N,77.267^{\circ}E)$ and SURF $(44.352^{\circ}N,103.751^{\circ}W)$.}
	\label{fig:detectorangle}
\end{figure}

We embed a spherical surface inside the Earth with radius $r_{\oplus}-d_{\rm lab}$, where $d_{\rm lab}$ is the depth at which the direct detection experiment of interest is placed underground. A detector with his fixed coordinates will travel through the isodetection rings in a non-trivial way, see appendix~\ref{ss:labpositionvelocity}. The position of a detector in terms of the isodetection angle $\Theta$ is given by
\begin{align}
	\Theta(t) = \arccos\left[\frac{\vec{v}_{\oplus}(t)\cdot \vec{x}_{\rm lab}(t)}{v_{\oplus}(t)(r_{\oplus}-d_{\rm lab})}\right]\, ,\label{eq:thetat}
\end{align} 
where we have to use~\eqref{eq:vearth} and~\eqref{eq:labpos}. Note that $\vec{v}_{\oplus}(t)$ only changes marginally over a matter of days. We show the evolution of~\eqref{eq:thetat} for different laboratories around the globe in figure~\ref{fig:detectorangle}. It illustrates how different experiments move through the isodetection rings and already hints at, which experiments will be more sensitive to the Earth's shadowing effect and the corresponding diurnal modulations.

More concretely, we divide up the detector sphere into 180 isodetection rings with width $\Delta \Theta = 1^{\circ}$ symmetric about the axis defined by $\vec{v}_{\oplus}(t)$. We label the rings with $\Theta_k$ where $k\in[0,179]$. The area of ring $\Theta_k$ is then
\begin{align}
	A_k = 2\pi (r_{\oplus}-d_{\rm lab})^2 \left[ \cos (\Theta_k) -\cos(\Theta_k+\Delta\Theta)\right]\, . \label{eq:ringarea}
\end{align}
We choose $\Delta\Theta$ sufficiently small, such that the velocity distribution will be approximately constant over a single isodetection ring's surface. The final goal of our MC simulations is to derive a precise estimate for the local speed distribution function for each of these rings based on the individual simulated particles including statistical uncertainties.

For each simulated trajectory we record if, how often, where (in terms of isodetection ring $\Theta_k$), and with what velocity the particle passes through the detection sphere. The different isodetection rings have varying surface areas, see~\eqref{eq:ringarea}. Furthermore the particle flux will also vary with $\Theta$ because of the DM wind. As a consequence more particles pass through certain rings than others. Yet we want to have the same statistics for each ring. We therefore repeat simulating DM trajectories until we have collected the same velocity data sample size $N_{\rm sample}$ for each ring. This value must of course be chosen sufficiently large. This way we accumulate $180\times N_{\rm sample}$ DM velocity vectors. For each isodetection ring we can determine the local distribution functions, DM densities and event rates independently.

In order to get an accurate estimate for the speed distribution function we need a non-parametric density estimation method. We employ histograms, weighting the data properly, in order to estimate the distribution function. For details on the distribution estimates we refer to appendix~\ref{a:DD}. There we also show in detail how direct detection rates can be derived from the MC data. In this work we take a CRESST-II like detector as an illustrative example. Our framework can however easily be extended to other detector types or detection strategies.

Early MC simulations of a similar kind have been performed in the 90s~\cite{Collar1992,collar1993,Hasenbalg1997}. The focus of these works laid on the standard WIMP model with DM masses of 50 GeV and higher and direct detection using nuclear recoils, whereas we focus on sub-GeV DM and have new detection techniques in mind. Apart from this there is a number of essential differences in the implementation of \textsc{DaMaSCUS}, two of which we emphasize here. For one we employ the corrected method of finding appropriate initial conditions for the DM particles, as has been described above. This ensures a spatially uniform distribution inside the Earth and avoids density overestimation close to the surface. Secondly we embedded the isodetection rings underground instead of directly on the Earth's surface, so we can simulate cases, where the rock above an experiment acts as a DM shield. Naturally a very strong interaction between DM and nuclei is required for this to occur. Another minor difference is a more refined modelling of the Earth.

\section{Results}
\label{s:results}
This study investigates the terrestrial effect of DM-nucleus underground scatterings in the regime of sub-GeV mass and strong enough (yet unconstrained) DM-nucleus cross sections that allow single or multiple scatterings before the DM particle arrives at the detector. We start by presenting simulation results in the case where DM can scatter at most once, because in this case we can directly compare the MC results to the analytic ones obtained in~\cite{Kavanagh2016} with the \textsc{EarthShadow} code. As we will demonstrate our MC results agree perfectly with the ones of ~\cite{Kavanagh2016}, reinforcing our
confidence in our simulation code as we use it afterwards in cases of multiple DM scattering where the \textsc{EarthShadow} code is no longer valid. 

We choose four benchmark points for a DM mass of 500MeV, which lies directly at the boundary of the detection sensitivity of CRESST-II. The lowest cross-section used corresponds to an underground scattering probability of $10\%$. The MC simulation for this point will be compared directly to the analytic results of~\cite{Kavanagh2016}, given that a $10\%$ probability results at most to a single scattering (double scattering has a probability of $\sim 1\%$). The other three cross-sections are higher and tuned such that the average number of underground scatterings is 1, 10 and 50 respectively. For a summary see table~\ref{tab:benchmark}.

\begin{table}[h!]
\centering
	\begin{tabular}{|l|c|c|c|}
	\hline
		Simulation ID	&$\sigma^{\rm SI}_{\chi n}(0)$[pb]	&$\langle N_{\rm sc}\rangle$	&$N_{\rm sample}$	\\
		\hline
		\hline
		`SS'		&0.521								&0.12							&$10^7$		\\
		\hline
		`MS1'		&4.26								&1.0							&$10^7$		\\
		\hline
		`MS10'		&41.2								&10.0							&$10^6$\\
		\hline
		`MS50'			&300.0							&$\gtrsim$50.0						&$5\cdot10^5$	\\
		\hline
	\end{tabular}
\caption{Benchmark points for the MC simulations with $m_{\chi}=500$ MeV. Here $\langle N_{\rm sc}\rangle$ is the average number of underground scatterings of the simulated trajectories and the sample size $N_{\rm sample}$ is the number of recorded velocity data points per isodetection ring.}
\label{tab:benchmark}
\end{table}
The resulting DM density and velocity distribution functions can be used to compute direct detection event rates for any kind of experiment and search strategy.

The computations were performed on the Abacus 2.0, a 14.016 core supercomputer of the DeIC National HPC Center at the University of Southern Denmark and typically involve the simulation of up to $10^{11}$ DM particle trajectories.

\subsection{Single Scattering: Comparison to \textsc{EarthShadow}}
\label{ss:earthshadow}
\begin{figure}[tbp]
\includegraphics[width=0.49\textwidth]{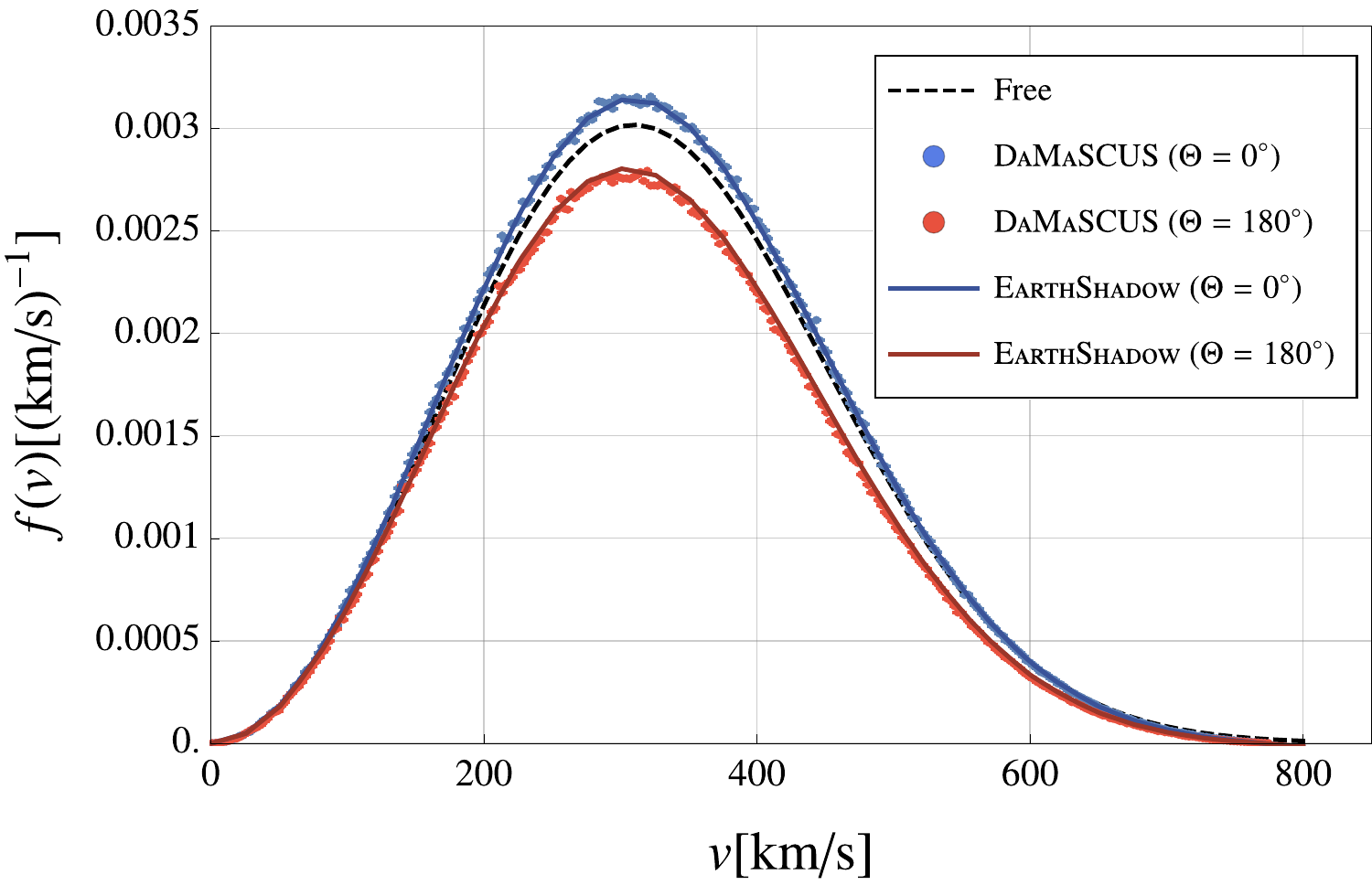}
\includegraphics[width=0.49\textwidth]{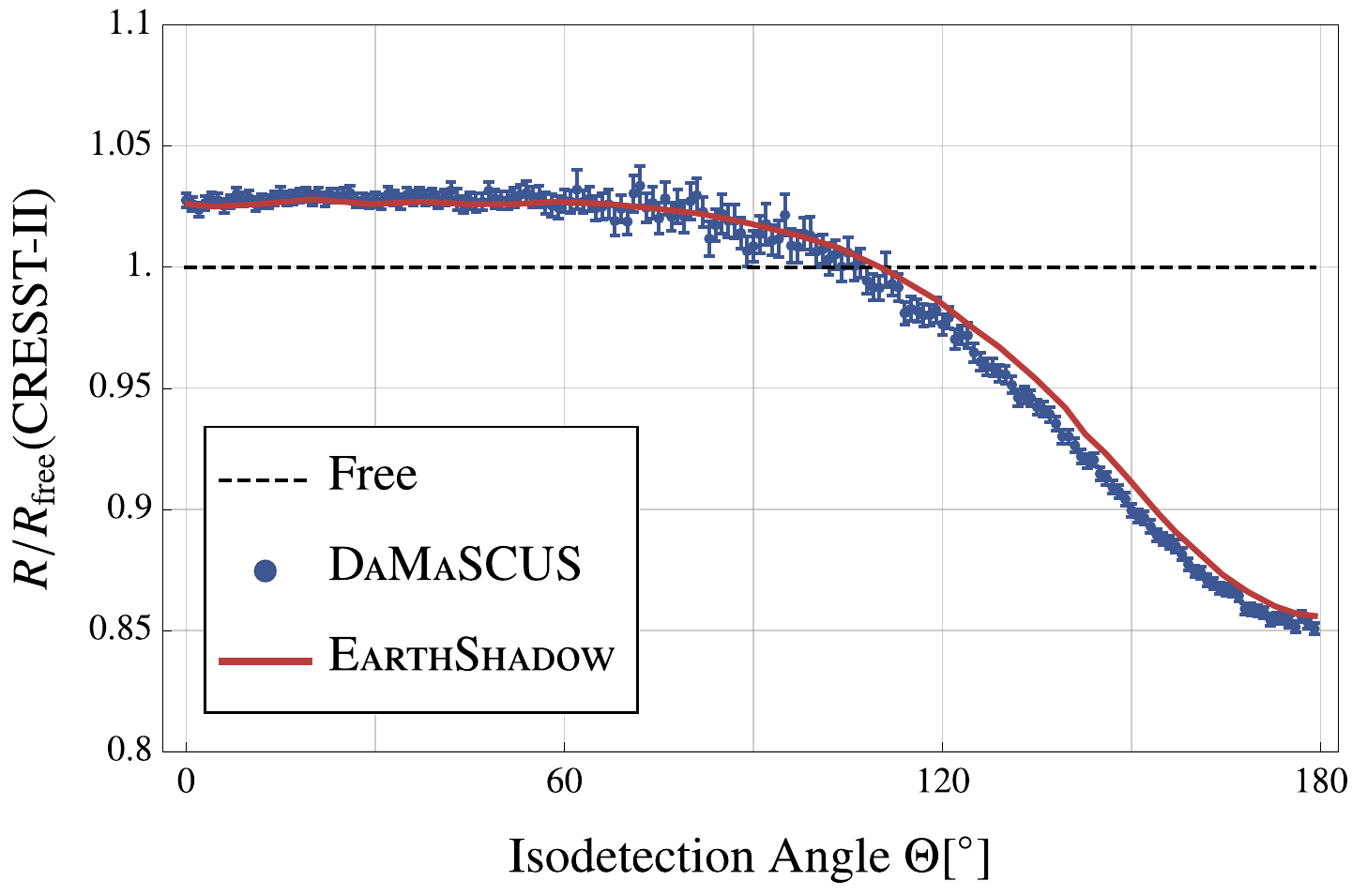}
	\caption{The left hand side shows a comparison between the analytic results of the \textsc{EarthShadow} code and our MC results for the local DM speed distribution at two different values of $\Theta$, as well as the normalized unperturbed halo distribution, which corresponds to a DM density of $0.3\text{GeV}/\text{cm}^3$. The others are accordingly normalized to the DM density for the respective isodetection ring. The right plot shows the event rate variation at a CRESST-II type detector over the globe for both the \textsc{EarthShadow} and the \textsc{DaMaSCUS} results. The analytic result is taken from the left panel of figure 7 in~\cite{Kavanagh2016}.}
	\label{abb:esresults}
\end{figure}
In order to perform a vital consistency check, we compare the results of the analytic methods of the \textsc{EarthShadow} code with our MC simulation. Since the analytic approach can only cover the single-scattering regime, we choose a DM-nucleon cross-section corresponding to an underground scattering probability of $10\%$. We can employ the respective \textsc{EarthShadow} routine and find $\sigma_{\chi n}^{\rm SI}(0)\approx 0.5$pb for a DM mass of 500 MeV. For these parameters we find good agreement with the MC simulations, where $\sim 90\%$ of the simulated particles cross the Earth freely, while the remaining scatter at least once.

As mentioned earlier, when comparing the two approaches we should keep in mind that deviations of the order of a percent should be expected, since about a percent of the particles will scatter not just once but twice, which is only accounted for by \textsc{DaMaSCUS}. These small deviations will affect the large $\Theta$ region, where DM particles have travelled the longest distances underground.

The distortions of the DM distribution result from deceleration and deflections of DM particles due to underground scatterings, which redistribute the DM inside the Earth. In the left panel of figure~\ref{abb:esresults} we see the speed distribution functions for $\Theta=0^{\circ}$ and $180^{\circ}$, which are normalized to a DM density of $0.3\text{ GeV cm}^{-3}$. Already in the single-scattering regime it is clear that for high values of $\Theta$ the DM density gets reduced, because DM particles travel longer distances underground and scatter away from their original path more likely. Particles entering the Earth at low values of $\Theta$ on the other hand have to cross only short underground distances of the order of the detector depth $d_{\rm det}$ and will most likely not scatter before reaching this depth. However other particles which originally were not on a path towards this region may be deflected and still end up here, leading to an overall increase in the local DM density. Overall the distortion of the DM distribution is due to deflections rather than the deceleration of DM.

The main outcome of this comparison however is the excellent agreement and consistency of the local DM distribution between \textsc{DaMaSCUS} and \textsc{EarthShadow}. Therefore it doesn't surprise that this agreement also translates to the direct detection event rate shown in the right panel of figure~\ref{abb:esresults}. It clearly shows the Earth's `shadow', the decrease of the signal rate for large values of $\Theta$. The decrease is mostly caused by the locally depleted DM population due to deflections. The effect of DM deceleration is less severe in this case, but might still be crucial for detection strategies relying on DM particles from the tail of the velocity distribution.

 Detectors, whose revolution around the Earth involve high isodetection angles would be able to measure this decrease as a diurnal modulation. As seen in figure~\ref{fig:detectorangle} this is a signature pronounced more in laboratories of the southern hemisphere. The experimental facilities beneath Gran Sasso for example only cover roughly $\Theta\in(0^{\circ},90^{\circ})$, where the signal rate is slightely increased but effectively constant. 
 
 One should mention here that there are two more sources of diurnal modulation in the DM signal that can compete or exceed the shadowing effect in low enough DM-nucleus cross sections. One is related to the rotational velocity of the Earth around its own axis ($\sim 1\text{km}/\text{s}$), which is superimposed with the total velocity of the Earth in the rest frame of the Galaxy, causing a small daily fluctuation of the DM flux of the order $\sim 10^{-4}$ at the detector. The second source is due to gravitational focusing, i.e., the fact that the Earth works as a gravitational lens causing small fluctuations on the DM density depending on the relative position of the detector with the center of the Earth and the direction of the DM wind~\cite{Kouvaris2015}. All three sources of modulation have different amplitudes and phases, producing a final modulation which is the superposition of all. 

\subsection{Multiple Scatterings}
\label{ss:multiple}
\begin{figure}[tbp]
	\centering
	\includegraphics[width=0.45\textwidth]{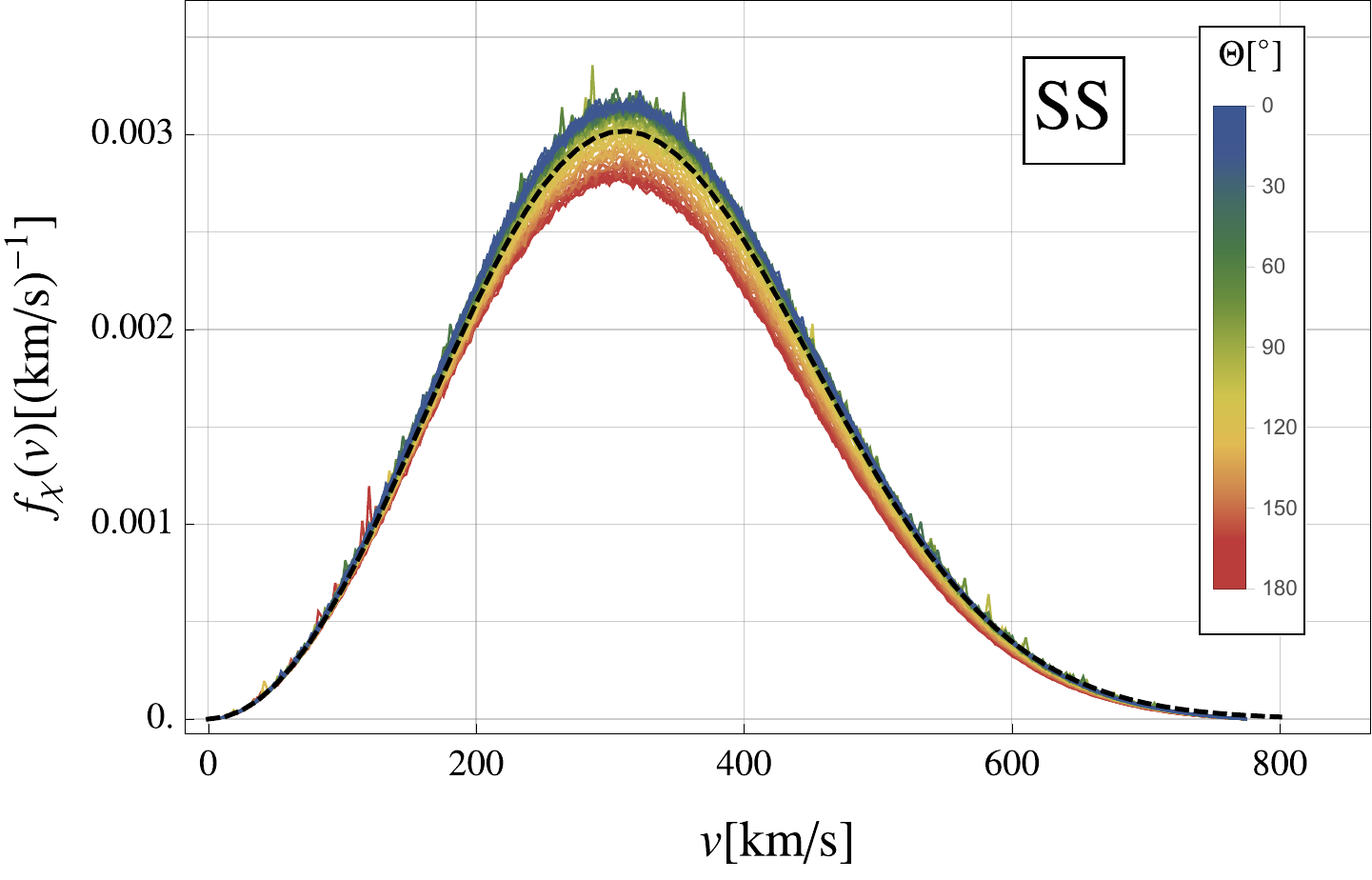}
	\includegraphics[width=0.45\textwidth]{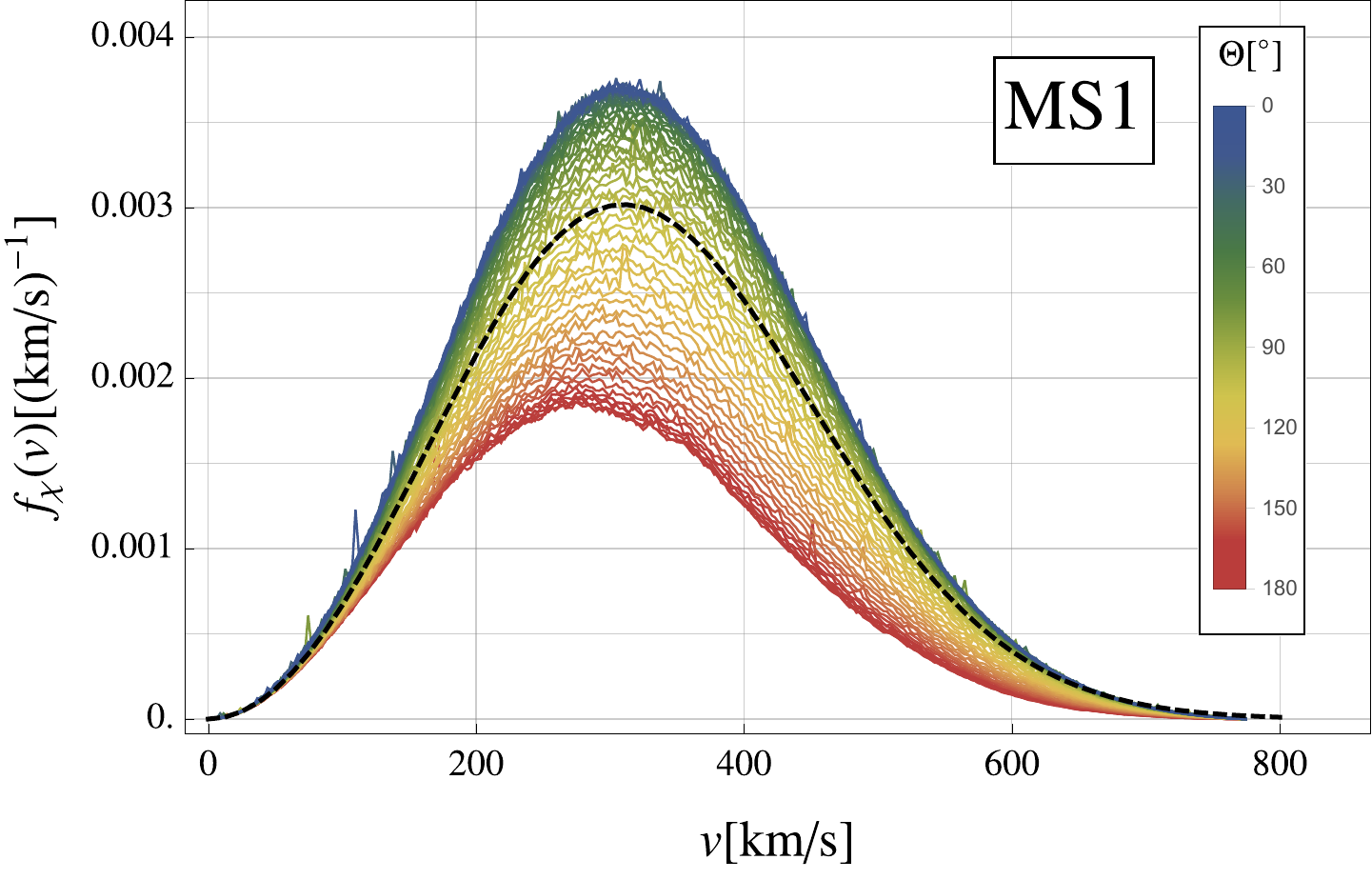}
	\includegraphics[width=0.45\textwidth]{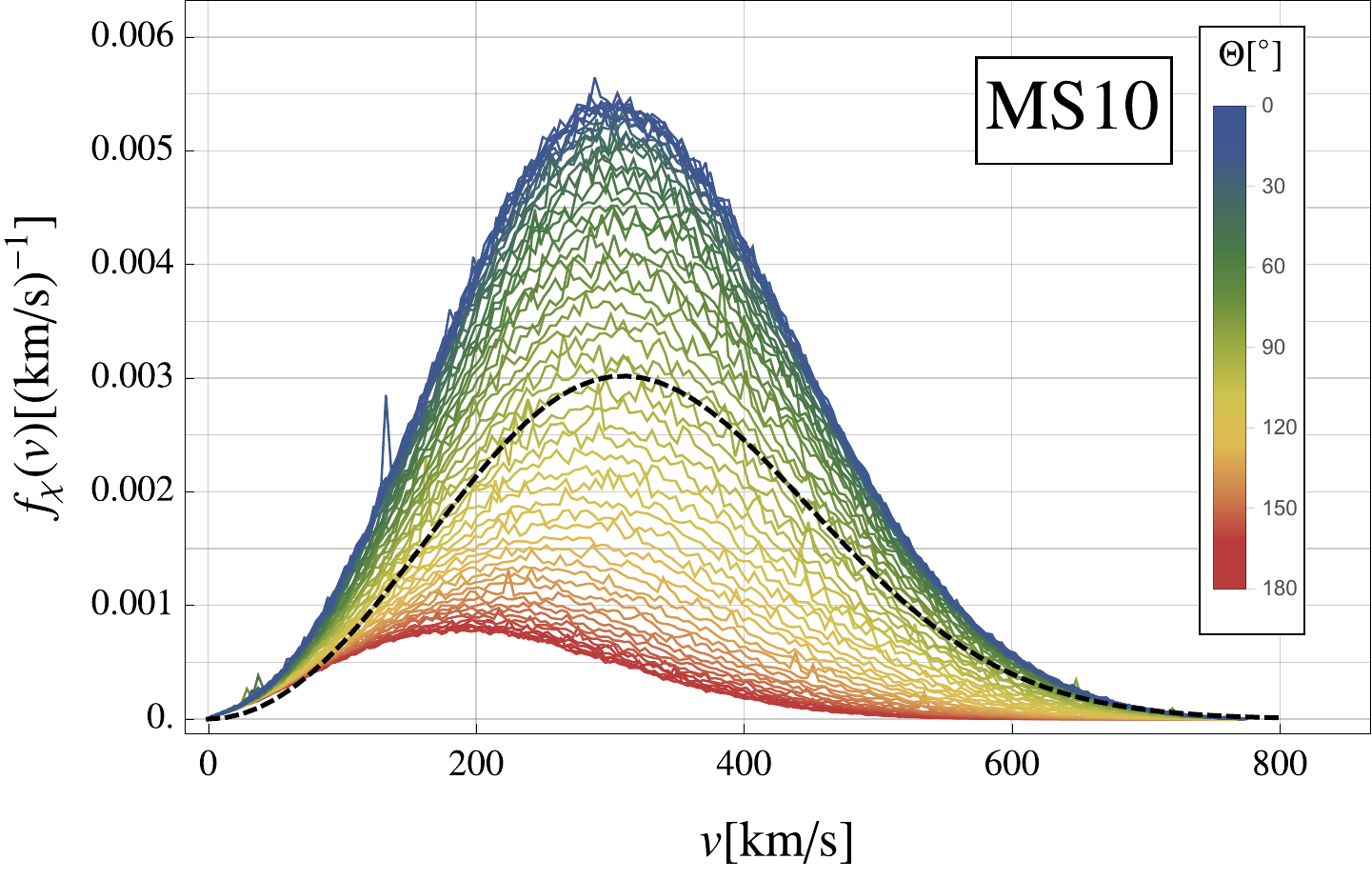}
	\includegraphics[width=0.45\textwidth]{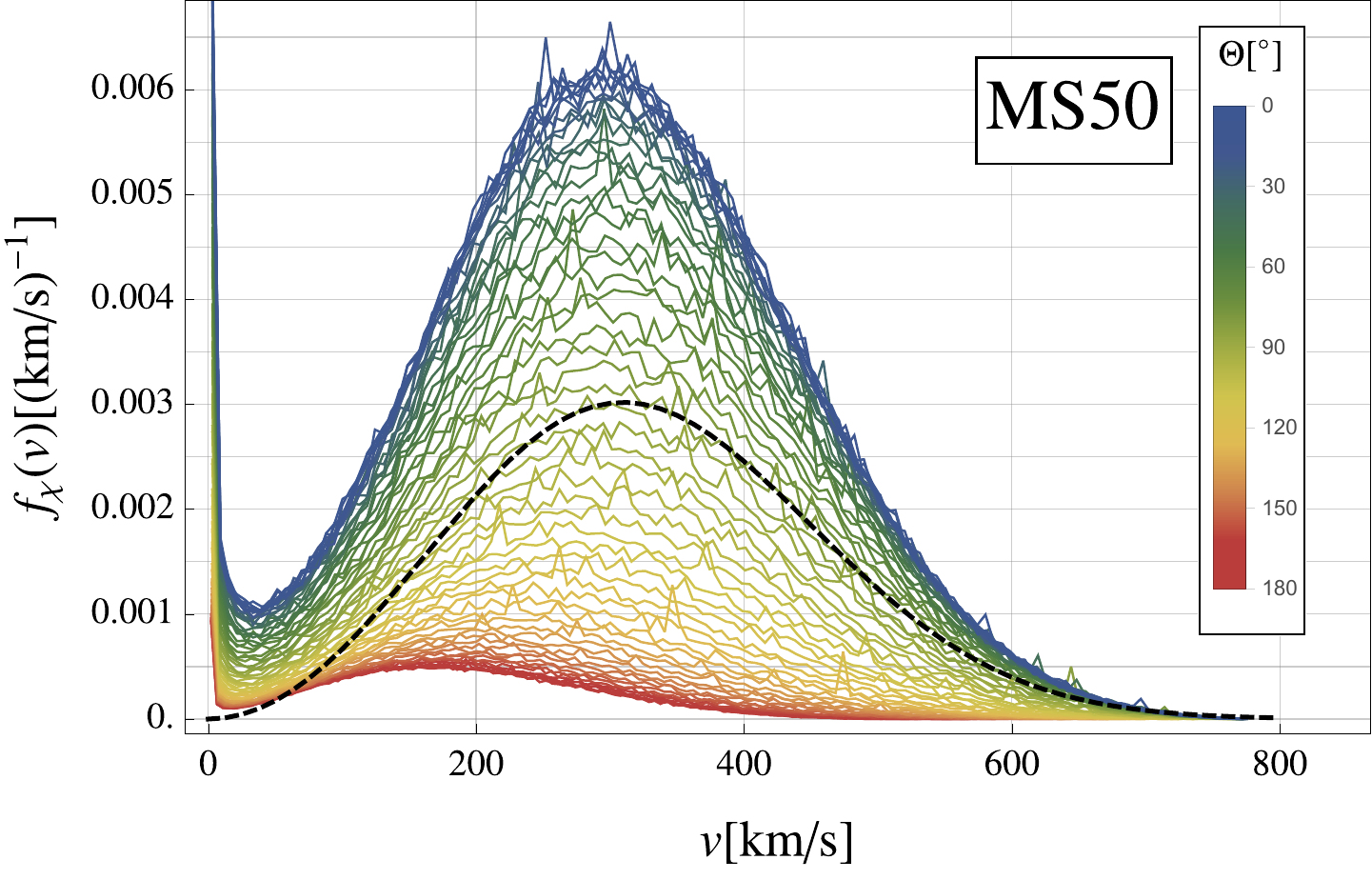}
	\caption{DM speed distributions across the globe for our four benchmark points. Note that they are normalized to $0.3$ GeV cm${}^{-3}$. The black dashed line shows the speed distribution of free DM.}
	\label{fig:globalphasespace}
\end{figure}
Now that consistency between the analytic and the MC results is established in the single-scattering regime we can confidently turn to higher cross sections of the multiple-scatterings or diffusion regime. We investigate three benchmark points, i.e., cross sections tuned to result in 1, 10 and 50 underground scatterings on average, see table~\ref{tab:benchmark}.

We start off with the local DM speed distribution distortions depicted in figure~\ref{fig:globalphasespace}. Compared to the single-scattering regime results, the two main observations are the much more severe deceleration and depletion of the DM population for higher values of $\Theta$, i.e. deep in the Earth's shadow. Especially for the benchmark point `MS50' we see clearly how underground scatterings deplete regions with high speeds and increase the slow DM population significantly. The effect is less pronounced in the other benchmark points, but it can be inferred by observing a horizontal shift in the peak of the distributions as $\Theta$ changes. On the other hand the overall height of the DM speed distributions functions show how the deflections of DM particles enhance the DM density for locations facing the DM wind, while simultaneously deplete the DM population for large values of $\Theta$. We again see the same effects as in the single-scattering regime but in a more severe way. A new feature for cross sections of the order of $\mathcal{O}(100\text{pb})$ is the second peak in the lower right panel on figure~\ref{fig:globalphasespace}, populated by very slow particles. Although current detector thresholds cannot probe this second peak, more sensitive future detectors could observe this peak as a bump in the low energy recoil spectrum. Something similar has been claimed to be produced by DM gravitationally bound to the Earth~\cite{Catena2016b,Catena2016a}.

\begin{figure}[tbp]
 \centering
\includegraphics[width=0.45\textwidth]{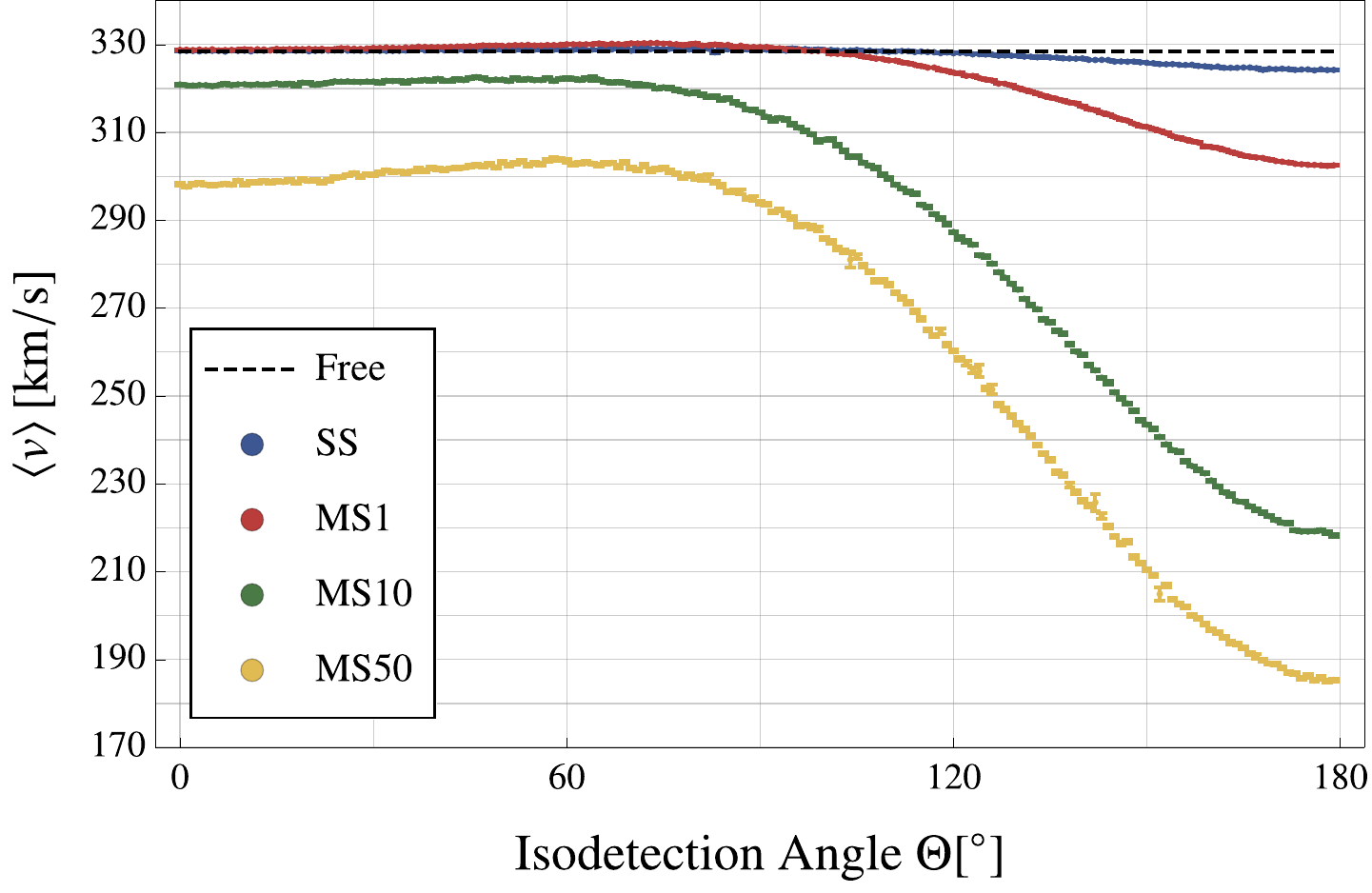}
\includegraphics[width=0.45\textwidth]{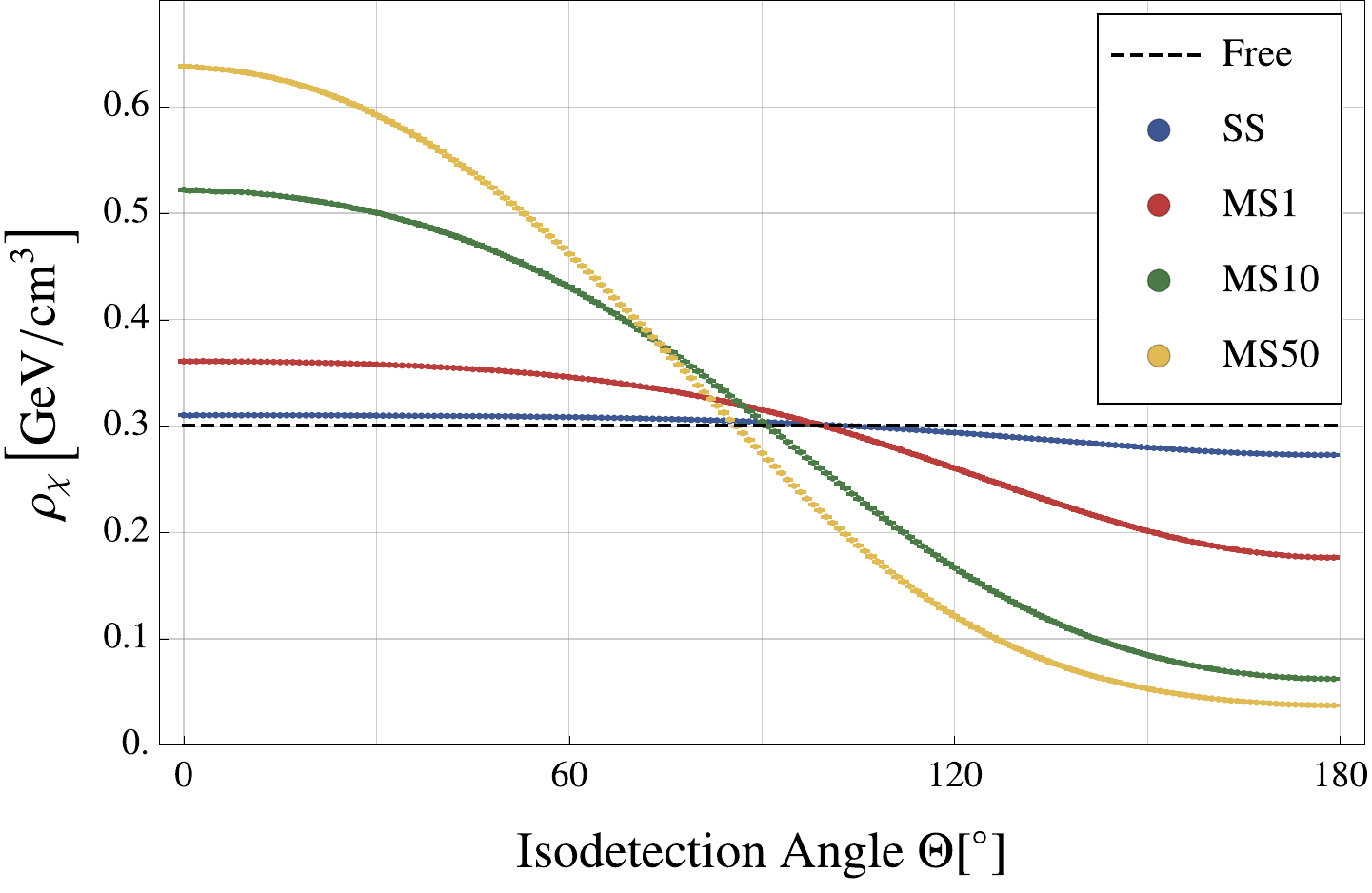}
\includegraphics[width=0.45\textwidth]{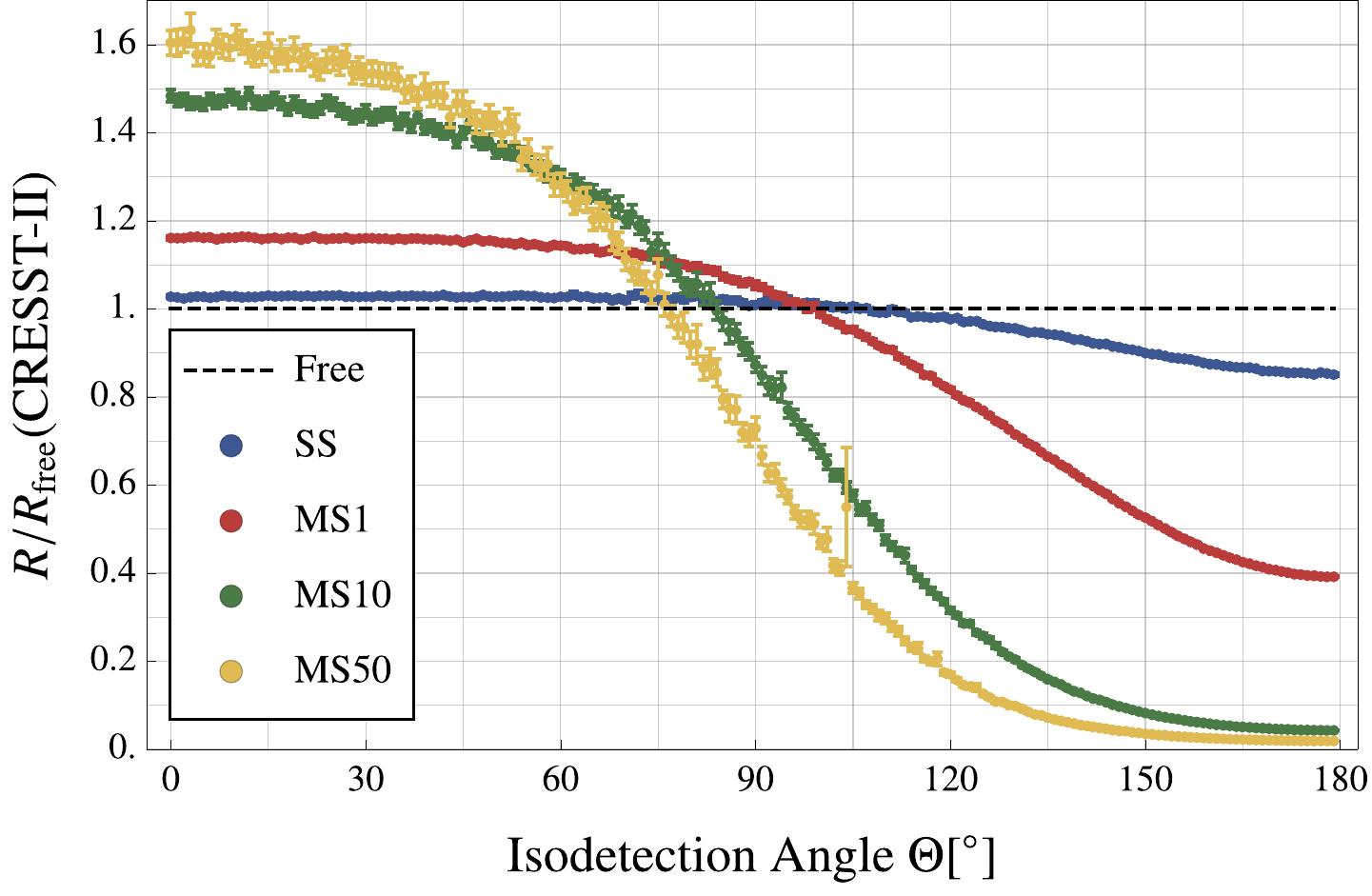}
\includegraphics[width=0.45\textwidth]{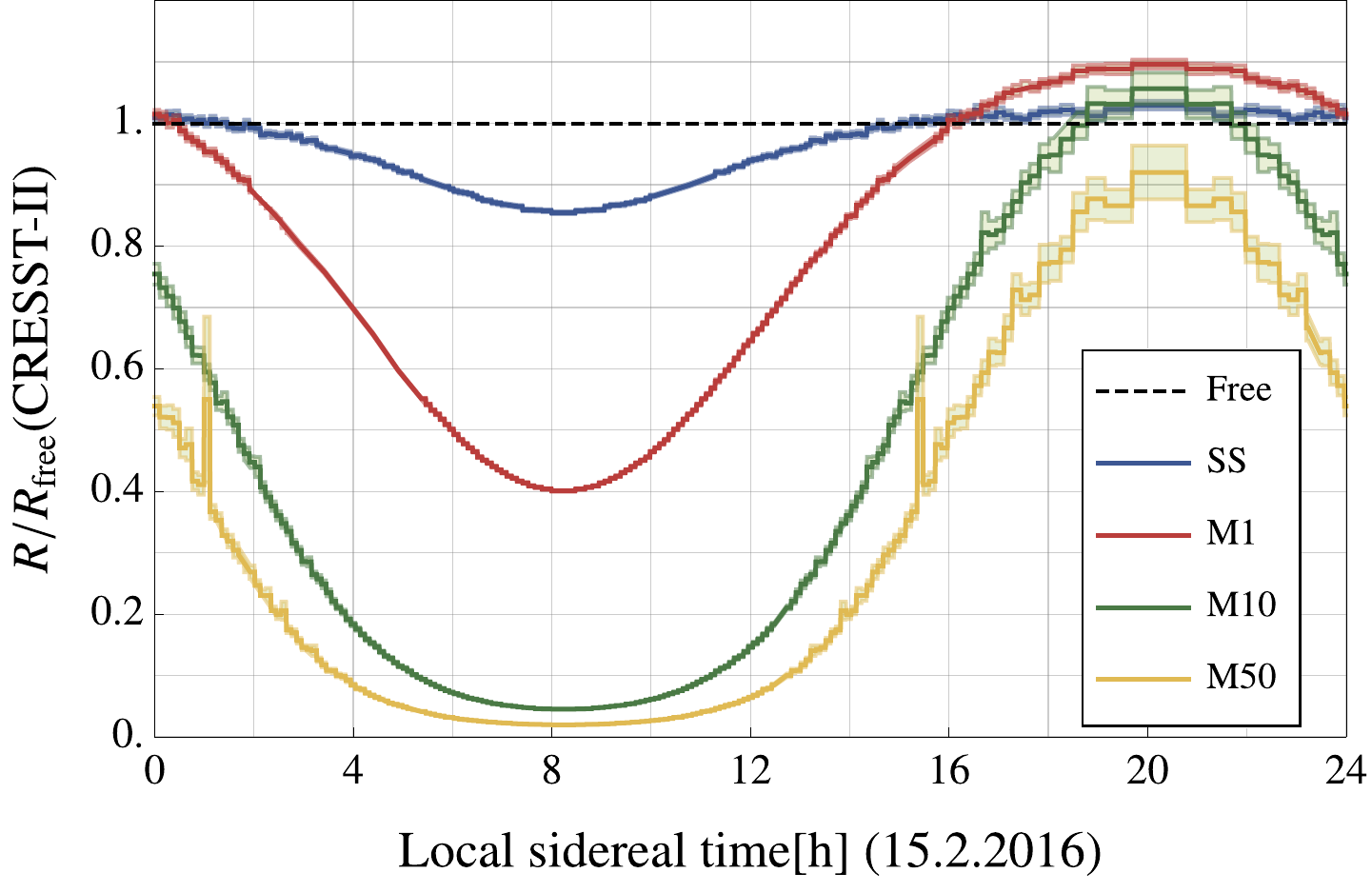}
\caption{\textsc{DaMaSCUS} results: The top left panel shows the local average DM speed as a function of the isodetection angle $\Theta$. The particles reaching high values of $\Theta$ are notably slowed down through scatterings. The top right panel shows the decline of the local DM density as we move deeper into the Earth shadow and more particles get deflected, a fact that is also clear from figure~\ref{fig:globalphasespace}. The two lower panels show the event rate for a CRESST-II type detector. The left shows the local signal rate for any value of $\Theta$. As an actual experiment revolves around the Earth axis it moves through the isodetection rings following Eq.~\eqref{eq:thetat}, resulting in diurnal modulation of the direct detection signal rate. The lower right panel shows this effect during a random day of the year for a hypothetical experiment at the SUPL in the southern hemisphere ($37.07^{\circ}$S), where these modulations have the largest amplitudes.}
\label{abb:multi2}
\end{figure}

The local DM velocity distribution functions in figure~\ref{fig:globalphasespace} are the central output of \textsc{DaMaSCUS}, since they allow us to compute direct detection rates of any kind. They also encode the local average speed and DM density, which we plot separately in the first two panels of figure~\ref{abb:multi2} for the four benchmark points. Depending on the interaction strength the average speed is decreased through nuclear stopping with a clear minimum for large $\Theta$. We observe a slight but noticeable increase for the isodetection rings between 60 and $90^{\circ}$. It is most notable for the `MS50' benchmark point, but it is also visible in `MS1', where we see that the average speed is increased slightly even beyond the expectation for free particles. It could be explained by particles from the fast velocity tail of our initial distribution which typically enter the Earth parallel to the DM wind at low values of $\Theta$ and deflect towards high-$\Theta$ isodetection rings.

In addition another impact of the elastic DM-nucleus collisions is the redistribution of DM particles. The local DM density, while being increased by up to 100\% at $\Theta=0^{\circ}$ for the benchmark point with the strongest interaction, drops to values below the halo density as we move further into the Earth's DM shadow. The decrease of both speed and density results to a decrease of direct detection events, as we see in the left lower panel of figure~\ref{abb:multi2}, where we again took a CRESST-II like detector as a
n example. The event rate drops significantly more than in the single-scattering regime. Therefore the overall local DM density is higher in the northern hemisphere compared to the southern, with an increasing difference as we move to stronger DM-nucleon interactions. 

\begin{figure}[tbp]
\centering
\includegraphics[width=0.6\textwidth]{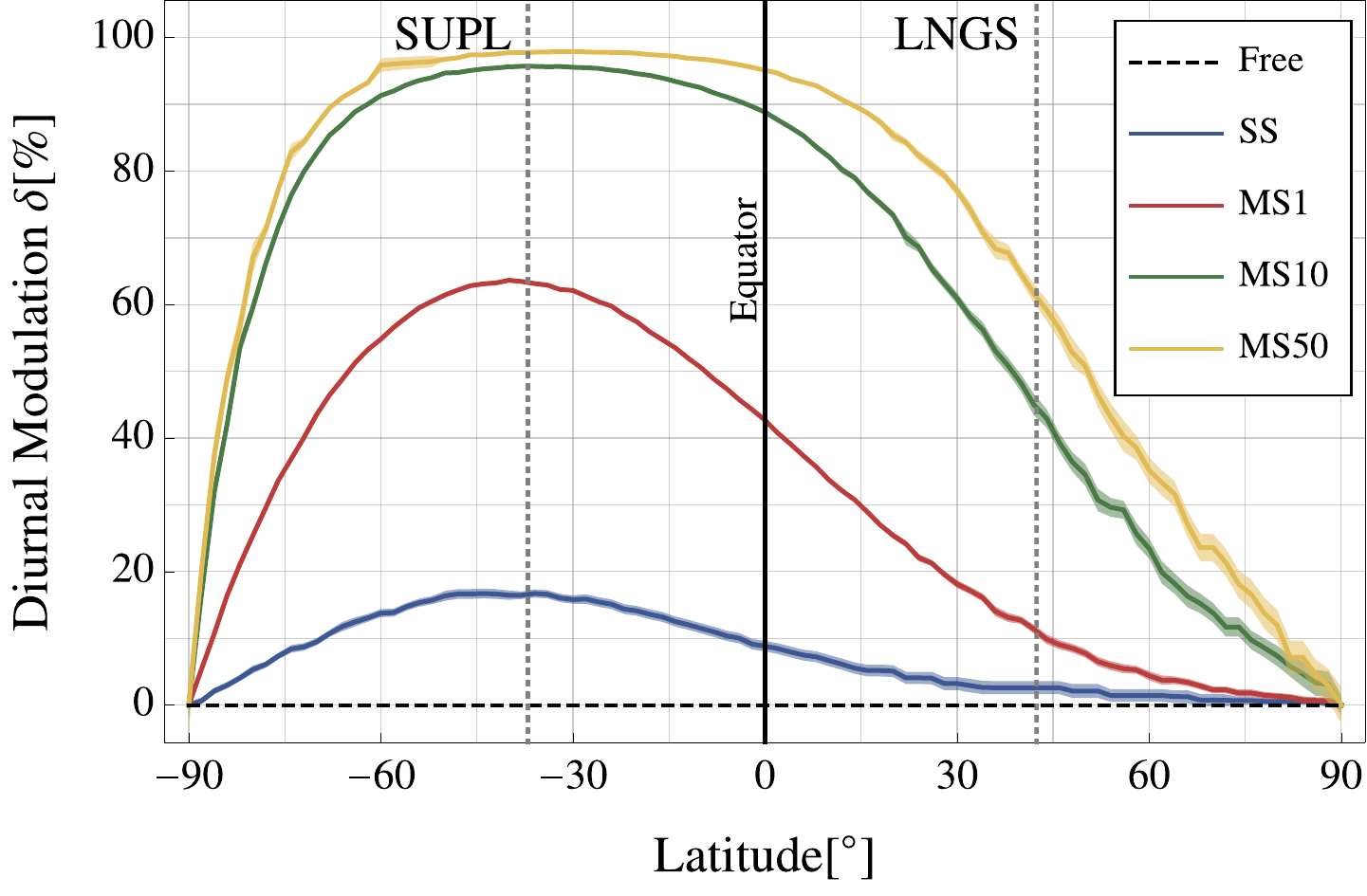}
\caption{Diurnal event modulation as a function of the experiment's latitude for the four benchmark points. The grey dotted lines indicate the LNGS in the northern and the SUPL in the southern hemisphere.}
\label{fig:globalmodulation}
\end{figure}
The thereby induced diurnal modulation of the signal rate for a detector of fixed location on Earth traversing the isodetection rings in accordance with Eq.~\eqref{eq:thetat} is shown in the lower right panel of figure~\ref{abb:multi2}, where we selected the SUPL ($37.07^{\circ}$S,$142.81^{\circ}$E) detector in Australia as its location in the southern hemisphere is very sensitive to diurnal modulations of this kind. However such modulations are not a signature reserved only for experiments in the South. To see the dependence on an experiment's latitude, we define the percentile signal modulation $\delta$ observable at a given laboratory via 
\begin{align}
	\delta(\Phi_{\rm lab}) =100\; \frac{R_{\rm max}-R_{\rm min}}{R_{\rm max}}\;\%\; ,
\end{align}
where $\Phi_{\rm lab}$ is the laboratory's latitude. We show the modulation as a function of the latitude in figure~\ref{fig:globalmodulation} for the four benchmark points.

It is generally true that the diurnal modulation caused by underground scatterings is maximal for experiments in the southern hemisphere. In the multiple-scattering regime however we see that such modulations can be significant almost anywhere on the globe with the exception of the poles' neighbourhood. Even for experiments at Gran Sasso we find a diurnal signal modulation of 10\%, 45\% and almost 60\% for $\langle N_{\rm sc}\rangle=1$, 10 and 50 respectively. Nevertheless a location such as the SUPL is strongly favoured with modulations of $\sim 18\%$, $\sim 65\%$ and more than $95 \%$ for the same set of cross sections. For the two benchmark points with the largest cross-section we find that a potential sub-GeV direct detection experiment in the southern hemisphere should expect a vastly reduced signal rate during a significant part of the day while the planet's bulk mass shields off the sought particles. In this case we have significant sensitivity to DM during certain hours of the day only.

\section{Conclusions and Outlook}
\label{s:conclusion}
In this paper we study the effect of underground DM-nucleus scatterings on the direct detection prospects of sub-GeV DM with sufficiently large cross sections that could accommodate such pre-detector underground scatterings. This effect can be important in various contexts. One is the possibility of a subdominant strongly interacting component of DM not abiding by direct detection constraints. Additionally, DM in the sub-GeV region is practically unconstrained by current direct detection underground experiments. Since in this region, DM can have sufficiently strong DM-nucleus interactions, underground scatterings before DM particles reach the detector must be taken into account. At best, these underground scatterings can distort the nuclear recoil spectrum making hard to establish beyond any doubt DM discovery. In the worst case scenario, detectors at current deep sites might be completely blind to this part of DM parameter space regardless of exposure and even if they lower significantly their energy thresholds. Similarly for DM constraints based on DM-electron interactions in the same mass region, one should consider carefully the terrestrial effect from nuclear stopping before reaching the detector.

In order to study this effect in full generality, we developed the \textsc{DaMaSCUS} code where we can simulate the trajectory of halo DM particles of given cross section and mass that cross the Earth, potentially scatter underground and eventually scatter in the detector. In order to do this, we use a state-of-the-art density and composition profile of the Earth and improve the generation of initial conditions from simulations done in the past. The output of our simulation is the local DM density and velocity distribution of DM at the location of a given underground detector, time and day of a year.

For DM-nucleon cross section where DM particles can scatter at most once before reaching the detector, the results of our MC simulation are in excellent agreement with the analytical calculation of ~\cite{Kavanagh2016}. However our simulations have the big advantage that they can be used for higher yet experimentally allowed DM-nucleon cross sections, where DM particles can possibly scatter more than once. In fact we presented several benchmark points where the average number of underground scatterings is 1, 10 and 50. As expected, we found that with increasing cross section the local DM density and velocity distribution deviate more and more from the transparent Earth, where the effect of underground scatterings is ignored completely. We found that this affects strongly the number of potential events in an underground detector. This is due to two effects: DM deceleration and DM deflection due to underground scatterings. We found that the effect can be so strong, that a potential DM signal in that parameter space will have a large diurnal modulation simply because as the Earth rotates around its own axis, DM particles travel different distances in order to reach the detector, thus increasing the probability/number of underground scatterings. We predicted the amount of diurnal modulation for different labs and we verified that the southern hemisphere has larger daily fluctuations in the DM signal. Note that for sufficiently strong cross section, the diurnal modulation can be next to $\sim 100\%$. Our simulation can also provide real time correlations between DM signals of detectors in different latitudes, facilitating the task of discriminating DM from potentially other backgrounds with daily modulation. Our study revealed also another interesting feature. For sufficiently large cross section, the velocity distribution acquires a second sharp peak at low energies. Although this is currently out of experimental reach, with the advent of detectors with lower thresholds, this peak could create a very distinct feature in the low energy recoil spectrum, identifying DM beyond any doubt. 

We leave a lot of things for future work. We plan to extend our work by studying the shadowing effect for different types of DM-nuclei interactions beside the spin-independent one as well as for DM-electron interactions. The latter is extremely interesting since the core of the Earth is assumed to be metallic. This means that one should treat some of the atomic electrons as free particles. It should be stressed here that free electrons do not require a minimum energy to excite. They can absorb even tiny amounts of energy, thus decelerating even low energetic DM particles. In addition we intend to investigate the shadowing effect in the case of long range forces between DM and nuclei, e.g. in models where the interaction is mediated by a light dark photon. In this case DM particles might interact collectively with multiple atoms as they pass through, losing energy perhaps in a similar fashion as massive objects lose energy through gravitational dynamical friction. We also plan to include in our simulation the other two sources of diurnal modulation, i.e., due to gravitational focusing and the rotational velocity of the Earth. Moreover we plan to study the shadowing effect on directional detectors. A first attempt was presented in~\cite{Kouvaris2016}. Finally we plan to use the simulation in order to make a precise estimate of the effect of DM gravitationally bound to the Earth on the nuclear recoil spectrum of underground detectors. Although there is a promising analytical estimate~\cite{Catena2016b,Catena2016a} that can be explored by future detectors, a precise MC simulation will give a more accurate estimate about the density and velocity distribution of bound DM in the Earth.

\begin{acknowledgments}
This work is partially funded by the Danish National Research Foundation, grant number DNRF90 and by the Danish Council for Independent Research, grant number DFF – 4181-00055. Computation/simulation for the work described in this paper was supported by the DeIC National HPC Centre, SDU.
\end{acknowledgments}

\appendix

\section{Astronomical Conventions and Coordinates}
\label{a:astro}
For the sake of completeness and in order to serve as a reference we review the astronomical basics required for our simulations. 
\subsection{Sidereal Time}
In order to keep track of the Earth's rotation, we use a time unit based on that very rotation called sidereal time. A sidereal day is the time interval of one rotation relative to vernal equinox $\Upsilon$. It is slightly shorter than a mean solar day: $23.9344699$ hours~\cite{almanac2014}. Note that the sidereal time is often given as an angle instead of a time unit.

The local apparent sidereal time (LAST) is the time since the local meridian passed $\Upsilon$. In this section we will show how to compute the LAST for any time and location. But first, we have to introduce a reference time. The time in our problem are always given relative to 01.01.2000 12:00 TT (or GMT), a commonly used reference time denoted as J2000.0. In order to calculate the fractional number of days $n_{\text{J2000.0}}$ relative to J2000.0 for a given date $D.M.Y$ and time $h:m:s$ (UT) we use the following relation~\cite{McCabe2013},
\begin{align}
	n_{\text{J2000.0}} &= \lfloor 365.25\tilde{Y}\rfloor +\lfloor 30.61(\tilde{M}+1)\rfloor+D\nonumber\\
	&\quad+\frac{h}{24}+\frac{m}{24\times 60}+\frac{s}{24\times 60^2} -730563.5\, ,
	\intertext{where}
	\tilde{Y} &= \begin{cases}
		Y-1\quad &\text{if } M=1\text{ or }2\, ,\\
		Y\quad &\text{if }M>2
	\end{cases}\, ,\\
	\tilde{M} &= \begin{cases}
		M+12\quad &\text{if } M=1\text{ or }2\, ,\\
		M\quad &\text{if }M>2
	\end{cases}\, ,
\end{align}
and $\lfloor \cdot \rfloor$ is the floor function. For example the 15.02.2016, 8:00am corresponds to $n_{\text{J2000.0}} = 3332.83$. We also define the epoch,
\begin{align}
	T_{\text{J2000.0}}\equiv\frac{n_{\text{J2000.0}}}{36525}\, .
\end{align}
Next we show in detail how to compute the Local Apparent Sidereal Time (LAST) anywhere on Earth starting from the Universal Time. We start with the formula for the Greenwich mean sidereal time (GMST) in seconds~\cite{almanac2014},
\begin{align}
	\text{GMST} &= 86\,400s\,\big[0.7790\,5727\,32640+n_{\text{J2000.0}}\text{mod } 1+0.0027\,3781\,1911\,35448\;n_{\text{J2000.0}}\big]\nonumber\\
	+&0.000\,967\,07s+307.477\,102\,27s\,T_{\text{J2000.0}}+0.092\,772\,113s\,T_{\text{J2000.0}}^2+\mathcal{O}(T_{\text{J2000.0}}^3)\, .	\label{eq:gmst}
\end{align}
For the Greenwich apparent sidereal time (GAST) we'll have to add the equation of equinoxes,
\begin{align}
	\text{GAST} &= \text{GMST} + E_e(T_{\text{J2000.0}})\, ,
	\intertext{for which we'll use the following approximation.}
	E_e(T_{\text{J2000.0}})&\approx\Delta\psi\cos\epsilon_A + 0.000176s \sin \Omega+ 0.000004s\sin 2\Omega.
	\intertext{Here we have}
	\Delta\psi&\approx-1.1484s\,\sin \Omega - 0.0864s\, \cos 2L\, ,\nonumber\\
	\Omega &= 125.0445\,5501^{\circ}-0.0529\,5376^{\circ}n_{\text{J2000.0}} + \mathcal{O}(T_{\text{J2000.0}}^2)\, ,\nonumber\\
	L &= 280.47^{\circ}-0.98565^{\circ}n_{\text{J2000.0}} + \mathcal{O}(T_{\text{J2000.0}}^2)\, ,\nonumber\\
	\epsilon_A &=23.4392\;79444^{\circ}-0.01301021361^{\circ}T_{\text{J2000.0}} + \mathcal{O}(T_{\text{J2000.0}}^2)	\nonumber\, .
\end{align}
To obtain the local apparent sidereal time (LAST) at a location with latitude and longitude $(\Phi,\lambda)$, we just add the longitude,
\begin{align}
	\text{LAST}(\lambda) = \text{GAST} + \frac{\lambda}{360^{\circ}}86400s\, .
\end{align}
Note that for western longitudes, $\lambda$ is negative. It should always be made sure that $\text{LAST}\in (0,86400)$. Comparing to the tables of~\cite{almanac2014}, the errors of these approximations are of the order of tens of milliseconds.

\subsection{Coordinate Systems}
\label{ss:coordinatesystems}
\begin{figure}[tbp]
    \begin{minipage}[b]{0.5\textwidth}
    \centering
	\includegraphics[width=\textwidth]{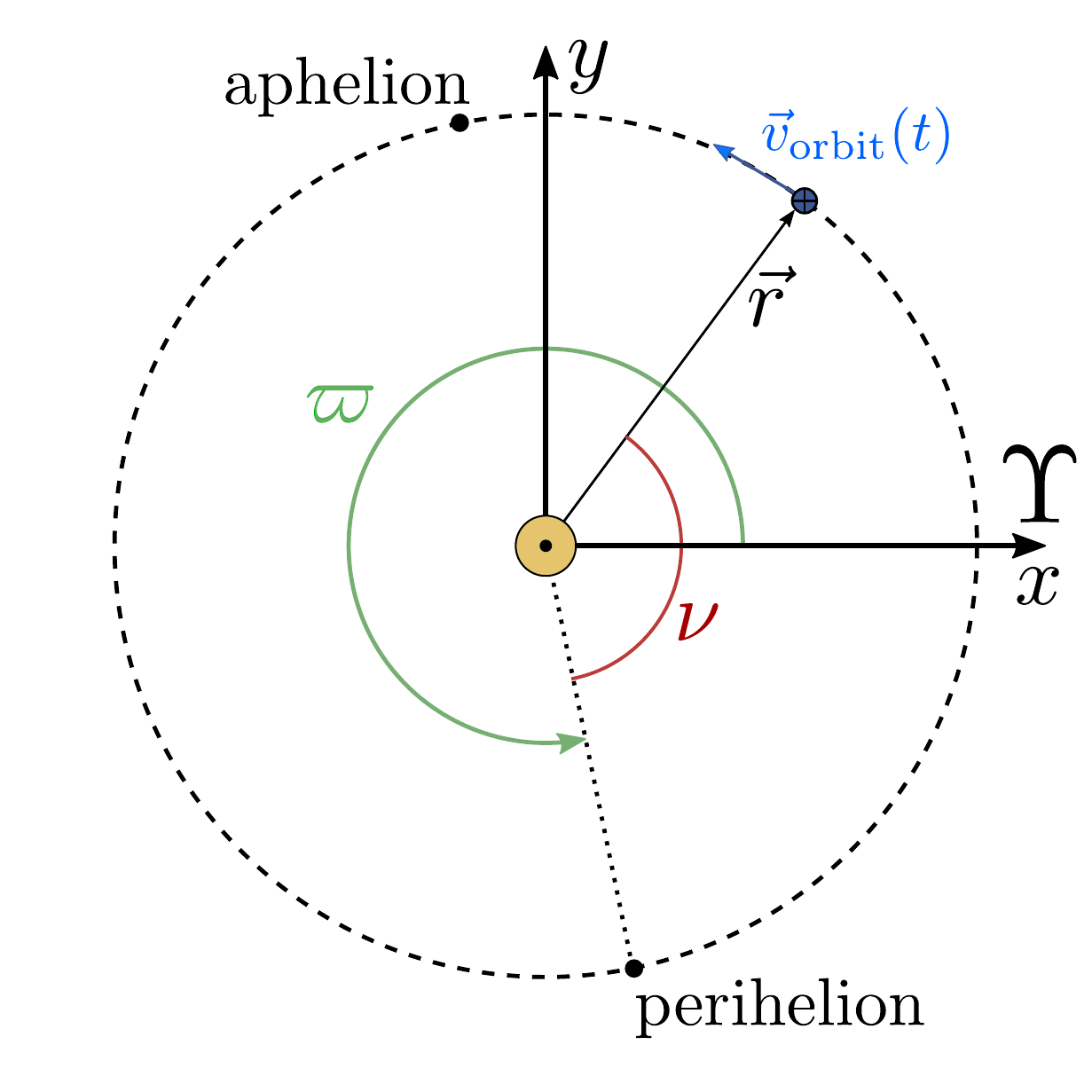}
	\caption{The Earth's orbital velocity in heliocentric ecliptic coordinates}
	\label{fig:heliocentriccoordinates}
    \end{minipage}
    \hfill
    \begin{minipage}[b]{0.5\textwidth}
    \centering
    \includegraphics[width=\textwidth]{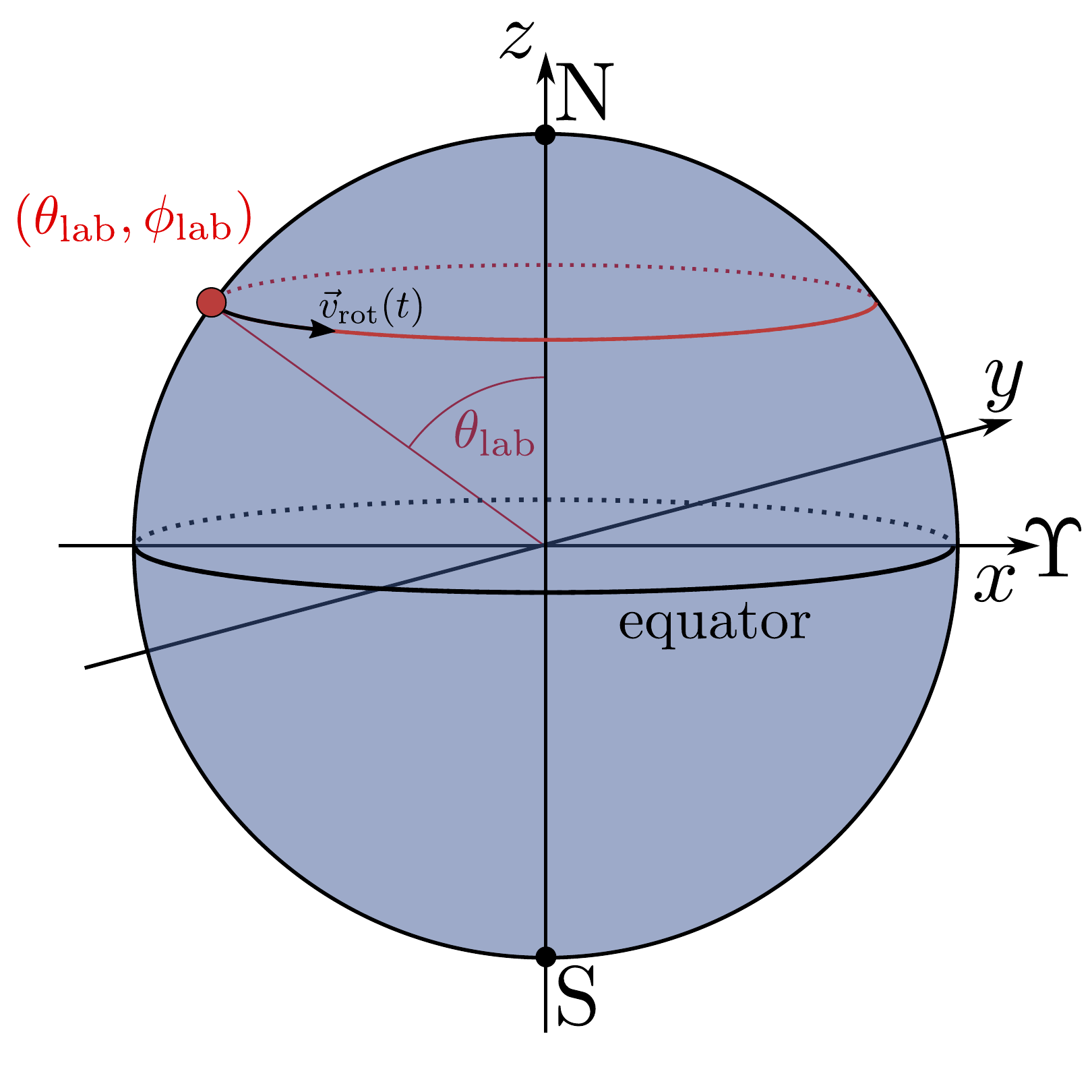}
	\caption{Laboratory position and velocity in equatorial coordinates}
	\label{fig:equatorialcoordinates}
    \end{minipage}
\end{figure}

We introduce the relevant coordinate systems and how to transform in between them~\cite{McCabe2013,almanac2014}. All coordinate systems are rectangular and right-handed.
\begin{enumerate}
	\item the galactic frame `(gal)': A heliocentric coordinate system, the $x$-axis points towards the galactic center, the $z$-axis points towards the galactic north pole. The $x$- and $y$-axis span the galactic plane.
	\item the heliocentric, ecliptic frame `(hel-ecl.)': A heliocentric coordinate system, the $x$-axis points towards vernal equinox $\Upsilon$, the $z$-axis points towards the ecliptic north pole. The $x$- and $y$-axis span the ecliptic plane. See figure~\ref{fig:heliocentriccoordinates}.
	\item the geocentric, ecliptic frame `(geo-ecl.)': A geocentric coordinate system, the $x$-axis points towards vernal equinox $\Upsilon$, the $z$-axis points towards the ecliptic north pole. The $x$- and $y$-axis span the ecliptic plane.
	\item the geocentric, equatorial frame `(equat)': A geocentric coordinate system, the $x$-axis points towards vernal equinox $\Upsilon$, the $z$-axis points towards the Earth north pole. The $x$- and $y$-axis span the equatorial plane. See figure~\ref{fig:equatorialcoordinates}.
	\item the laboratory frame `(lab)': A coordinate system with the detector in the origin. The $x$-axis points towards east, the $y$-axis towards north and the $z$-axis to the sky.
\end{enumerate}
Since all our calculations will in the end be done in the galactic frame, we need the transformation matrices, which are time-dependent.

\paragraph{1.) (lab) $\longleftrightarrow$ (equat):} The rotation from the laboratory frame to the equatorial frame is done by
\begin{align}
	\vec{x}^{\text{(equat)}} = \mathcal{N}\vec{x}^{\text{(lab)}}\, ,\text{ with }
	\mathcal{N}=\begin{pmatrix}
		-\sin \phi		&-\cos \theta \cos \phi			&\sin \theta \cos \phi\\
		\cos \phi		&-\cos \theta \sin \phi			&\sin \theta \sin \phi\\
		0				&\sin \theta					&\cos \theta
	\end{pmatrix}\, ,
\end{align}
where $\theta = \frac{\pi}{2}-\Phi$ and $\phi = 2\pi\frac{\text{LAST}(\Phi,\lambda)}{86400\text{s}}$.

\paragraph{2.) (hel-ecl.)$\longleftrightarrow$(geo-ecl.):} The simplest transformation is the one between the two ecliptic frames,
\begin{align}
	\vec{x}^{\text{(geo-ecl)}}=-\vec{x}^{\text{(equat)}}\, .
\end{align}

\paragraph{3.) (geo-ecl.) $\longleftrightarrow$ (equat):}
 To transform a vector $\vec{x}^{\text{(geo-ecl)}}$ to equatorial coordinates, the necessary rotation is
\begin{align}
	\vec{x}^{\text{(equat)}} &= \mathcal{R}\vec{x}^{\text{(geo-ecl)}}\, ,\text{ with } \mathcal{R}=\begin{pmatrix} 1&0&0\\0&\cos \epsilon &-\sin \epsilon \\ 0 &\sin \epsilon & \cos \epsilon \end{pmatrix}\, , 
\end{align}
where $\epsilon = 23.4393^{\circ}-0.0130^{\circ}T_{\text{J2000.0}}$ is the obliquity or axial tilt of the ecliptic.

\paragraph{4.) (equat)$\leftrightarrow$(gal):}
The equatorial frame at J2000.0 can be related to the galactic frame,
\begin{align}
	\vec{x}^{\text{(gal)}} &= \mathcal{M}\vec{x}^{\text{(equat)}}(J2000.0)\, ,
	\intertext{with}
	\mathcal{M}_{11}&=-\sin l_{\text{CP}} \sin \alpha_{\text{GP}} - \cos l_{\text{CP}}\cos \alpha_{\text{GP}} \sin\delta_{\text{GP}} \, ,\nonumber\\
	\mathcal{M}_{12}&=\sin l_{\text{CP}} \cos \alpha_{\text{GP}}- \cos l_{\text{CP}}\sin \alpha_{\text{GP}}\sin \delta_{\text{GP}} \, ,\nonumber\\
	\mathcal{M}_{13}&=\cos l_{\text{CP}} \cos \delta_{\text{GP}} \, ,\nonumber\\
	\mathcal{M}_{21}&=\cos l_{\text{CP}} \sin \alpha_{\text{GP}}	-\sin l_{\text{CP}} \cos \alpha_{\text{GP}} \sin\delta_{\text{GP}} \, ,\nonumber\\
	\mathcal{M}_{22}&=-\cos l_{\text{CP}} \cos \alpha_{\text{GP}} -\sin l_{\text{CP}}\sin \alpha_{\text{GP}} \sin \delta_{\text{GP}} \, ,\nonumber\\
	\mathcal{M}_{23}&=\sin l_{\text{CP}} \cos\delta_{\text{GP}} \, ,\nonumber\\
	\mathcal{M}_{31}&=\cos \alpha_{\text{GP}} \cos\delta_{\text{GP}} \, ,\nonumber\\
	\mathcal{M}_{32}&=\sin \alpha_{\text{GP}}\cos\delta_{\text{GP}} \, ,\nonumber\\
	\mathcal{M}_{33}&= \sin	\delta_{\text{GP}}\, ,\nonumber
\end{align}
The three angles, namely the J2000.0 right ascension of the north galactic pole $\alpha_{\text{GP}}$, the J2000.0 declination of the north galactic pole $\delta_{\text{GP}}$ and the longitude of the north celestial pole in J2000.0 galactic coordinates $l_{\text{CP}}$, are
\begin{align}
	\alpha_{\text{GP}}=192.85948^{\circ}\, ,\quad	\delta_{\text{GP}}=27.12825^{\circ}\, ,\quad	l_{\text{CP}}=122.932^{\circ}\, .\nonumber
\end{align}
To rotate a vector in equatorial coordinates at J2000.0 to any time epoch $T_{\text{J2000.0}}$, we use
\begin{align}
	\vec{x}^{\text{(equat)}}(T_{\text{J2000.0}})&=\mathcal{P}\vec{x}^{\text{(equat)}}(J2000.0)\, ,
	\intertext{with}
	\mathcal{P}_{11}&=\cos \zeta_A \cos \theta_A \cos z_A - \sin\zeta_A \sin z_A \, ,\nonumber\\
	\mathcal{P}_{12}&=-\sin\zeta_A \cos\theta_A \cos z_A- \cos\zeta_A \sin z_A \, ,\nonumber\\
	\mathcal{P}_{13}&=-\sin\theta_A \cos z_A  \, ,\nonumber\\
	\mathcal{P}_{21}&=\cos\zeta_A \cos\theta_A \sin z_A+ \sin\zeta_A \cos z_A \, ,\nonumber\\
	\mathcal{P}_{22}&=-\sin\zeta_A \cos\theta_A \sin z_A+\cos\zeta_A \cos z_A \, ,\nonumber\\
	\mathcal{P}_{23}&=-\sin\theta_A \sin z_A \, ,\nonumber\\
	\mathcal{P}_{31}&=\cos\zeta_A \sin\theta_A \, ,\nonumber\\
	\mathcal{P}_{32}&=-\sin\zeta_A \sin\theta_A \, ,\nonumber\\
	\mathcal{P}_{33}&=\cos\theta_A \, .\nonumber
\end{align}
The equatorial precession angles are
\begin{align}
	\zeta_A &=2306.083227'' T_{\text{J2000.0}}+0.298850'' T_{\text{J2000.0}}^2\, ,\nonumber\\
	z_A&=2306.077181'' T_{\text{J2000.0}}+1.092735'' T_{\text{J2000.0}}^2\, ,\nonumber\\
	\theta_A &=2004.191903'' T_{\text{J2000.0}}+0.429493'' T_{\text{J2000.0}}^2\, .\nonumber
\end{align}

\paragraph{Summary:} In order to transform from any frame to any other frame, we just multiply the corresponding rotation matrices. We can follow this flow chart, where we take the inverse matrix if opposing the arrow's direction.
\begin{center}
	\begin{tikzpicture}[scale=0.7]
			\node[draw,rectangle] (a1) at (0.0,3.0) {(hel-ecl.)};
			\node[draw,rectangle] (b1) at (3.5,3.0) {(geo-ecl.)};
			\node[draw,rectangle,text width=1.4cm] (c1) at (7.0,3.0) {(equat)\\ at $T_{\text{J2000.0}}$};
			\node[draw,rectangle] (d1) at (10.0,3.0) {(lab)};
			\node[draw,rectangle,text width=1.4cm] (c2) at (7.0,0.0) {(equat)\\ at J2000.0};
			\node[draw,rectangle] (b2) at (3.5,0.0) {(gal)};	
			\draw[thick,<->] (a1.east)--(b1.west)node[pos=0.5,above]{$-\mathds{1}$};
			\draw[thick,->] (b1.east) to node[pos=0.5,above]{$\mathcal{R}$}(c1.west);
			\draw[thick,->] (d1.west) to node[pos=0.5,above]{$\mathcal{N}$}(c1.east);
			\draw[thick,->] (c2.north) to node[pos=0.5,right]{$\mathcal{P}$}(c1.south);
			\draw[thick,->] (c2.west) to node[pos=0.5,above]{$\mathcal{M}$}(b2.east);
	\end{tikzpicture}
\end{center}
For example, to go from equatorial to galactic coordinates for any given epoch $T_{\text{J2000.0}}$ we use
\begin{align}
	\vec{x}^{\text{(gal)}}&=\mathcal{M} \mathcal{P}^{-1}\vec{x}^{\text{(equat)}}(T_{\text{J2000.0}})\, ,
	\intertext{and to go from heliocentric, ecliptic coordinates to galactic coordinates we use}
	\vec{x}^{\text{(gal)}}&=-\mathcal{M} \mathcal{P}^{-1}\mathcal{R}\vec{x}^{(\text{hel-ecl)}}\, .
\end{align}
Now it is easy to write the axis vectors $\vec{e}_x=(1,0,0)^T$, $\vec{e}_y=(0,1,0)^T$ and $\vec{e}_z=(0,0,1)^T$ of the equatorial and heliocentric-ecliptic frame transformed into the galactic one,
\begin{align}
	\vec{e}_{x,\text{equat}}^{\text{(gal)}} &= \mathcal{M} \mathcal{P}^{-1}\vec{e}_x=\begin{pmatrix}
		-0.0548763\\0.494109\\-0.867666
	\end{pmatrix}
	+ \begin{pmatrix}
		0.0242316\\0.002688\\-1.546 \cdot 10^{-6}
	\end{pmatrix}T_{\text{J2000.0}} +\mathcal{O}(T_{\text{J2000.0}}^2)\, ,\label{eq:equat1}\\
	\vec{e}_{y,\text{equat}}^{\text{(gal)}} &= \mathcal{M} \mathcal{P}^{-1}\vec{e}_y=\begin{pmatrix}
		-0.873436\\-0.444831\\-0.198076
	\end{pmatrix}
	+ \begin{pmatrix}
		-0.001227\\0.011049 \\-0.019401
	\end{pmatrix}T_{\text{J2000.0}} +\mathcal{O}(T_{\text{J2000.0}}^2)\, ,\label{eq:equat2}\\
	\vec{e}_{z,\text{equat}}^{\text{(gal)}} &= \mathcal{M} \mathcal{P}^{-1}\vec{e}_z=\begin{pmatrix}
		-0.483836\\0.746982\\0.455984
	\end{pmatrix}
	+ \begin{pmatrix}
		-0.000533 \\0.004801\\-0.008431
	\end{pmatrix}T_{\text{J2000.0}} +\mathcal{O}(T_{\text{J2000.0}}^2)\, .\label{eq:equat3}
\end{align}
The axis vectors of the heliocentric-ecliptic frame are
\begin{align}
	\vec{e}_{x,\text{hel-ecl}}^{\text{(gal)}} &= -\mathcal{M}\mathcal{P}^{-1}\mathcal{R}\vec{e}_x=
	\begin{pmatrix}
	 0.054876\\-0.494109\\0.867666
	 \end{pmatrix} + 
	 \begin{pmatrix} -0.024232\\-0.002689\\1.546 \times 10^{-6}
	 \end{pmatrix}T_{\text{J2000.0}} +\mathcal{O}(T_{\text{J2000.0}}^2)\, ,\label{eq:helecl1}\\
	\vec{e}_{y,\text{hel-ecl}}^{\text{(gal)}} &= -\mathcal{M}\mathcal{P}^{-1}\mathcal{R}\vec{e}_y=\begin{pmatrix} 0.993821\\0.110992 \\0.000352 \end{pmatrix} + \begin{pmatrix} 0.001316\\-0.011851\\0.021267\end{pmatrix}T_{\text{J2000.0}} +\mathcal{O}(T_{\text{J2000.0}}^2)\, ,\label{eq:helecl2}\\
	\vec{e}_{z,\text{hel-ecl}}^{\text{(gal)}} &= -\mathcal{M}\mathcal{P}^{-1}\mathcal{R}\vec{e}_z=\begin{pmatrix} 0.096478\\-0.862286 \\-0.497147 \end{pmatrix} + \begin{pmatrix} 0.000227\\0.000015\\0.000018\end{pmatrix}T_{\text{J2000.0}} +\mathcal{O}(T_{\text{J2000.0}}^2)\, .\label{eq:helecl3}
\end{align}
This completes our review on coordinate systems.

\subsection{Earth's Velocity in the Galactic Frame}
\label{ss:earthvelocity}
The standard reference for the Earth velocity in the context of direct detection has long been the review by Smith and Lewin~\cite{Lewin1996}. However as pointed out first in~\cite{Lee2013} and confirmed in~\cite{McCabe2013} there is an error in the first order correction due to the orbit's eccentricity. We take the full expression for the Earth velocity vectors from~\cite{McCabe2013}. The Earth's velocity with respect to the galactic rest frame is the sum of three components,
\begin{align}
	\vec{v}_{\oplus}(t) = \vec{v}_r + \vec{v}_s + \vec{v}_e(t)\, .\label{eq:vearth}
\end{align}
These components are
\begin{enumerate}
	\item the galactic rotation,
	\begin{align}
		\vec{v}_{r} = \begin{pmatrix}
			0\\220\\0
		\end{pmatrix}\text{km s}^{-1}\, ,	
	\end{align}
	\item the sun's motion relative to nearby stars,
	\begin{align}
		\vec{v}_{s} = \begin{pmatrix}
		11.1\\12.2\\7.3
		\end{pmatrix}\text{km s}^{-1}\, ,	
	\end{align}
	\item the Earth's orbital velocity $\vec{v}_{e}(t)$ relative to the sun, visualized in figure~\ref{fig:heliocentriccoordinates},
	\begin{align}
	\vec{v}_{e}(t)&=- \langle v_{\oplus} \rangle\bigg[ \left(\sin L + e \sin (2L-\varpi)\right) \; \vec{e}_{x,\text{hel-ecl}}^{\text{(gal)}}\nonumber\\
	&\qquad\qquad+\left(\cos L + e \cos (2L-\varpi)\right) \; \vec{e}_{y,\text{hel-ecl}}^{\text{(gal)}}\bigg]\, ,
	\end{align}
	where the unit vectors $\vec{e}_{i,\text{hel-ecl}}^{\text{(gal)}}$ are given in Eqs.~\eqref{eq:helecl1} and~\eqref{eq:helecl2}.
\end{enumerate}
Finally we list the mean velocity $\langle v_{\oplus} \rangle$, the Earth's eccentricity $e$, the mean longitude $L$ as well as the longitude of the perihelion $\varpi$,
\begin{align}
	\langle v_{\oplus} \rangle &= 29.79\text{km s}^{-1} \, ,\quad	&e&=0.01671\, , \nonumber\\
	L&=\big[ 280.460^{\circ}+0.9856474^{\circ}n\big] \text{mod}\;360^{\circ}\, , \quad &\varpi &=\big[ 282.932^{\circ}+0.0000471^{\circ}n\big] \text{mod}\;360^{\circ}\, .\nonumber
	\end{align}
The inclusion of the correction due to the orbit's eccentricity may not be relevant for the results reported in this paper. We include them for the sake of completeness and potential future applications of the simulation code.

\subsection{Laboratory Position and Velocity}
\label{ss:labpositionvelocity}
As a first step we find the spherical coordinate angles $(\theta_{\text{lab}},\phi_{\text{lab}})$ of the detector's position in the geocentric equatorial coordinate system. We specify the location of a detector in the Earth through the latitude and longitude $(\Phi_{\text{lab}},\lambda_{\text{lab}})$ and the underground depth $d_{\text{lab}}$ of the laboratory.

The x-axis of the equatorial coordinate system points towards the vernal equinox or March equinox. Therefore the spherical coordinate angles are given by
\begin{align}
\theta_{\text{lab}}&= \frac{\pi}{2}-\Phi_{\text{lab}}\, ,\quad	\phi_{\text{lab}}(t)=\omega_{\text{rot}}\text{LAST}(\Phi_{\text{lab}},\lambda_{\text{lab}})\, .
\end{align} 
Since we use sidereal seconds, the rotation frequency is simply $\omega_{\text{rot}}=\frac{2\pi}{86400s}$. 
Now that we have the lab's spherical coordinates we can transform the position vector into galactic coordinates,
\begin{align}
	\vec{x}^{\text{(gal)}}_{\text{lab}} = \mathcal{M}\mathcal{P}^{-1}\begin{pmatrix*}[l]
	(r_{\oplus}-d_{\text{lab}})\sin \theta_{\text{lab}} \cos\phi_{\text{lab}}\\
	(r_{\oplus}-d_{\text{lab}})\sin \theta_{\text{lab}}\sin\phi_{\text{lab}}\\
	(r_{\oplus}-d_{\text{lab}})\cos\theta_{\text{lab}}
\end{pmatrix*}\, .\label{eq:labpos}
\end{align}
The velocity component of the laboratory due to Earth's rotation is given by
\begin{align}
	\vec{v}_{\text{rot}} &= \underbrace{\frac{2\pi r_{\oplus}}{T_d}}_{\equiv v_{\rm eq}}\cos \Phi_{\text{lab}} \mathcal{M}\mathcal{P}^{-1}\vec{e}^{\text{(equat)}}_{\phi}(\vec{x}_{\text{lab}})\nonumber\\
	&=-v_{\rm eq}\cos \Phi_{\text{lab}} \bigg( \sin(\phi_{\text{lab}}(t))\;\vec{e}_{x,\text{equat}}^{\text{(gal)}}-\cos(\phi_{\text{lab}}(t))\;\vec{e}_{y,\text{equat}}^{\text{(gal)}} \bigg)\, .
\end{align}
The unit vectors are given by~\eqref{eq:equat1} and~\eqref{eq:equat2}. The rotation velocity at the equator is $v_{\rm eq}\approx 0.465\text{ km s}^{-1}$.

\section{Earth Model}
\label{a:earth}
\begin{figure}[tbp]
	\begin{minipage}[b]{0.54\textwidth}
    	\centering
    	\resizebox{\textwidth}{!}{
		\begin{tabular}{lllllll}
		\hline
		$l$		&Layer		&Depth[km]		&$a_l$		&$b_l$		&$c_l$		&$d_l$		\\
		\hline
		0	&Inner Core			&0-1221.5		&13.0885		&0			&-8.8381		&0			\\
		1	&Outer Core			&1221.5-3480		&12.5815		&-1.2638		&-3.6426		&-5.5281	\\
		2	&Lower Mantle		&3480-5701		&7.9565		&-6.4761		&5.5283			&-3.0807	\\
		3	&Transition Zone I	&5701-5771		&5.3197		&-1.4836		&0				&0			\\
		4	&Transition Zone II	&5771-5971		&11.2494		&-8.0298		&0				&0			\\
		5	&Transition Zone III	&5971-6151		&7.1089		&-3.8045		&0				&0			\\
		6	&LVZ\& LID			&6151-6346.6		&2.6910		&0.6924		&0				&0			\\
		7	&Crust I			&6346.6-6356		&2.9			&0			&0				&0			\\
		8	&Crust II			&6356-6368		&2.6			&0			&0				&0			\\
		9	&Ocean				&6368-6371		&1.020		&0			&0				&0			\\
		10	&Space				& $>$6371			&0			&0			&0				&0			\\
		\hline
		\end{tabular}
		}
		\captionof{table}{Layer structure of the Preliminary Reference Earth Model (PREM) and its density profile coefficients~\cite{Dziewonski1981} given in $\text{g cm}^{-3}$. In the simulations we omit the ocean layer since experiments are located underground not underwater.}
		\label{tab:densitycoefficients}
    \end{minipage}
    \hfill
    \begin{minipage}[b]{0.46\textwidth}
    	\centering
		\includegraphics[width=\textwidth]{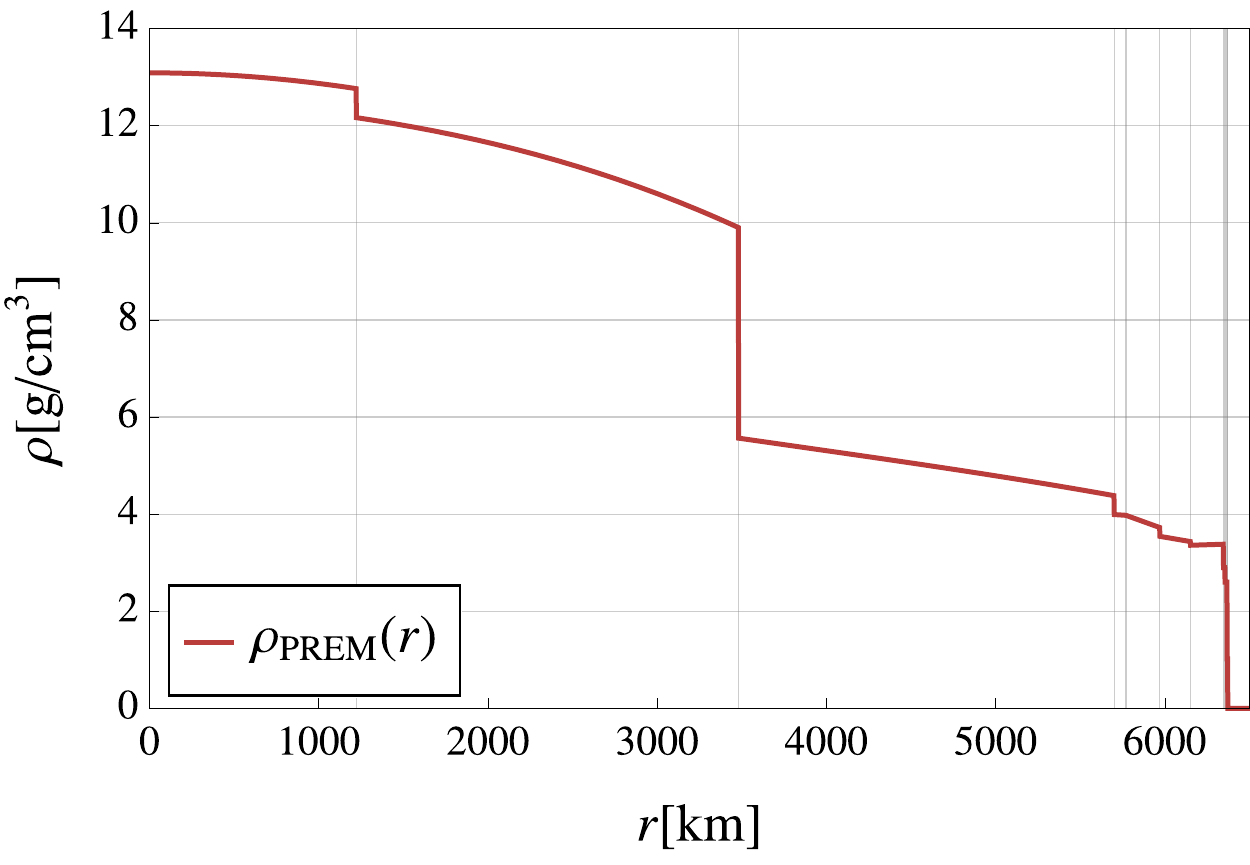}
		\caption{Density profile of the PREM.}
		\label{abb:PREMprofile}
    \end{minipage}
\end{figure}

We model planet Earth by dividing it into two sets of layers. On the one hand we have the compositional layers, which differ by their chemical element abundances. Based on~\cite{McDonough2013} we implement two compositional layers, the core and mantle. On the other hand we have ten mechanical layers, which, for our purpose, only differ by their density profile. We adopt the density profile of the Preliminary Reference Earth Model (PREM)~\cite{Dziewonski1981}. The chemical abundances of the two compositional layers are listed in table~\ref{tab:composition}. The core has a radius of about 3480km.
\begin{table}[tbp]
	\centering
	\begin{tabular}{|l|l|l|c|l|l|l|}
	\cline{1-3}\cline{5-7}
	Element&Core[\%] 		&Mantle[\%]	&&Element		&Core[\%] 		&Mantle[\%]			\\
	\cline{1-3}\cline{5-7}
	\isotope{Fe}{56}		&85.5			&6.26		&&\isotope{S}{32}		&1.9				&0.03	\\
	\isotope{O}{16}		&0				&44			&&\isotope{Cr}{52}			&0.9				&0.26	\\
	\isotope{Si}{28}	&6				&21			&&\isotope{Na}{23}			&0					&0.27	\\
	\isotope{Mg}{24}		&0				&22.8		&&\isotope{P}{31}		&0.2				&0.009	\\
	\isotope{Ni}{58}			&5.2				&0.2	&&\isotope{Mn}{55}		&0.3				&0.1	\\
	\isotope{Ca}{40}		&0				&2.53		&&\isotope{C}{12}		&0.2				&0.01	\\
	\isotope{Al}{27}		&0				&2.35		&&\isotope{H}{1}		&0.06				&0.01	\\
	\cline{1-3}\cline{5-7}
	\multicolumn{4}{c|}{}&\textbf{Total}		&\textbf{100.26}		&\textbf{99.83}			\\
	\cline{5-7}
	\end{tabular}
	\caption{Relative Element Abundances in the Earth Core and Mantle~\cite{McDonough2013}. For each element we use the most abundant isotope.}
	\label{tab:composition}
\end{table}
For the mechanical layers the Earth's density for each layer $l$ is parametrized as 
\begin{align}
\rho_{\oplus}(\vec{r}) = a_l +b_l x + c_l x^2 + d_l x^3 \, , \quad \text{where }x\equiv \frac{|\vec{r}|}{r_{\oplus}}\, . 	\label{eq:densityprofile}
\end{align}
The coefficients are given in~\cite{Dziewonski1981} and listed in table~\ref{tab:densitycoefficients}, the density profile is plotted in figure~\ref{abb:PREMprofile}.

\section{Hybrid-Algorithm for the Displacement Vector}
\label{a:displacement}
In this section we describe in detail how we solve Eq.~\eqref{eq:pxi} for $L$ using a mixture of analytic and numerical methods. First we can rewrite the equation as
\begin{align}
	\underbrace{\int\limits_{0}^{L/v}\dd t\;v\lambda_{\text{MFP}}^{-1}(\vec{x}(t),\vec{v})}_{\equiv \Lambda(L)} = -\log (1-\xi)\, . \label{eq:Lxi}
\end{align}
The straight forward approach to find $L$ and the displacement vector $\vec{\Delta}$ of a DM particle, e.g. in between two scatterings, is to solve~\eqref{eq:Lxi} by numerical integration along the particle's path. Since it is most likely that a particle, before it scatters or leaves the Earth, passes several layers, some maybe twice, without scattering, it will be reasonable to use the fact that we can solve the integral analytically inside a fixed layer $l$, as we will see now.

The first step towards the analytic solution for a given Earth layer $l$ and DM velocity $\vec{v}$ is to separate the space and velocity dependence of the mean free path in~\eqref{eq:lambdainverse},
\begin{align}
	\lambda_{\text{MFP}}^{-1}(\vec{x},\vec{v}) &= \rho_{\oplus}(\vec{x}) g_l(\vec{v})\, ,
\end{align}
where
\begin{align}
g_l(\vec{v}) &\equiv\sum\limits_i\frac{f_{A_i}}{m_{A_i}}\sigma_{\chi A}^{\rm total}(\vec{v})=\begin{cases}
g_{\text{core}}\;&0 \leq l<2\\g_{\text{mantle}}\;& 2 \leq l< 10
\end{cases}
\, .
\end{align}
The factors $g_l$ are constant inside each of the two compositional layers, since they only depend on the nuclear composition of the respective layer. As we have seen, in the case of SI interactions and light DM, we can neglect the nuclear form factor. In this case the factors $g_l$ lose their dependence on the velocity and only have to be computed once in the beginning of the MC simulations. Otherwise they have to be updated each time the DM particle changes its velocity.

 With the parametrisation of the density~\eqref{eq:densityprofile} we can solve the integral analytically for a constant layer $l$.
\begin{align}
	\Lambda_l(L)&\equiv\int\limits_{0}^{L/v}\dd t\;v\lambda_{\text{MFP}}^{-1}(\vec{x}(t)) =v g_l \int\limits_{0}^{L/v}\dd t\; \left( a_l +b_l \frac{|\vec{x}(t)|}{r_{\oplus}} + c_l \left(\frac{|\vec{x}(t)|}{r_{\oplus}}\right)^2 + d_l \left(\frac{|\vec{x}(t)|}{r_{\oplus}}\right)^3\right)\, , \nonumber\\
	&=\frac{g_l}{r_{\oplus}^2}\left(C_1 a_l + C_2 b_l + C_3 c_l + C_4 d_l\right)\, ,\label{eq:scatterprobabilityanalytic}
	\intertext{where $\vec{x}(t)=\vec{x}_0+t\vec{v}$. The coefficients are}
	C_1 &= L r_{\oplus}^2\, ,\nonumber\\
	C_2 &=\frac{r_{\oplus}}{2}\left(\tilde{L}(L+x_0 \cos \alpha)-x_0^2\cos \alpha+x_0^2\sin^2 \alpha  \log \left[ \frac{L+\tilde{L}+x_0 \cos \alpha}{x_0 (1+\cos \alpha)}\right]\right)\, ,\nonumber\\
	C_3 &=L\left(x_0^2+x_0 L \cos \alpha +\frac{1}{3}L^2\right)\, ,\nonumber\\
	C_4 &=\frac{1}{8r_{\oplus}}\bigg( (5-3\cos^2\alpha)(\tilde{L}- x_0)x_0^3\cos \alpha + 2 L^2 \tilde{L}(L+3x_o\cos\alpha)\nonumber\\
	&\qquad\qquad+L\tilde{L}x_0^2(5+\cos^2\alpha)+3x_0^4\sin^4\alpha\log \left[ \frac{L+\tilde{L}+x_0 \cos \alpha}{x_0 (1+\cos \alpha)}\right]\bigg)\, .\nonumber
\end{align}
We used $\cos \alpha = \frac{\vec{x}_0\cdot \vec{v}}{x_0v}$ and $\tilde{L}\equiv \sqrt{L^2+x_0^2+2Lx_0\cos\alpha}$.

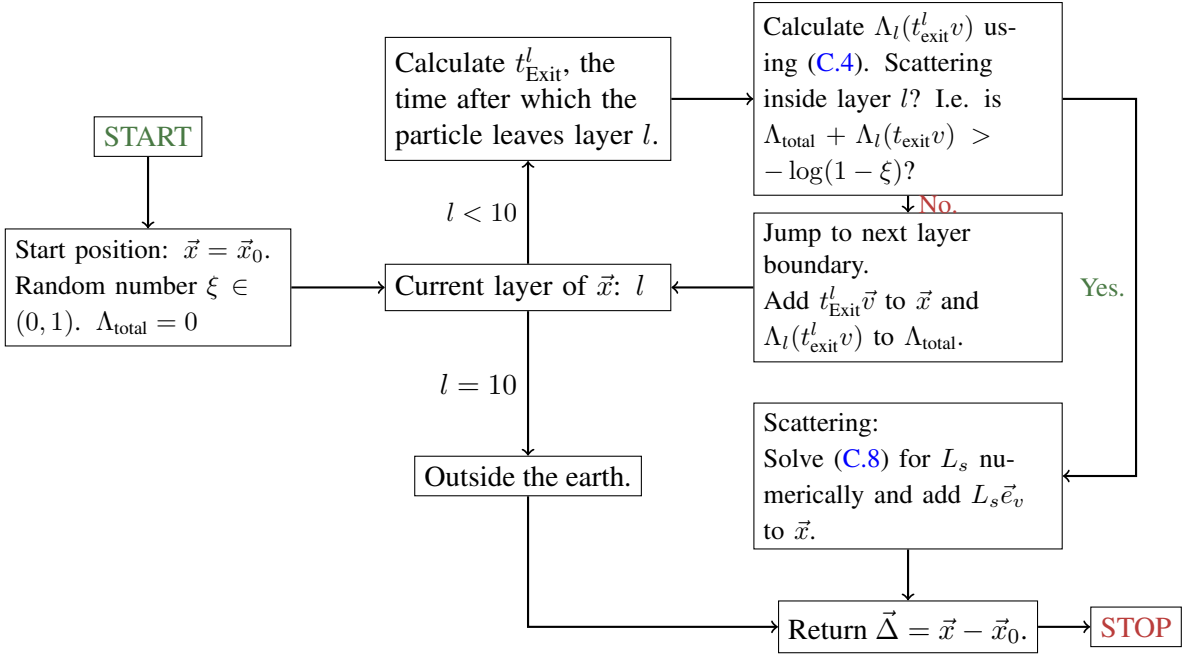
\begin{figure}[tbp]
	\centering
	\begin{tikzpicture}
		\node[draw,rectangle] (a0) at (0,0) {\color{myGreen}START};
		\node[draw,rectangle,text width=3.5 cm] (b0) at (0,-2) {\small Start position: $\vec{x}=\vec{x}_0$. Random number $\xi\in(0,1)$. $\Lambda_{\text{total}}=0$};
		\draw[thick,->](a0.south)--(b0.north);
		\node[draw,rectangle,text width=3.5cm] (b1) at (5,-2) {Current layer of $\vec{x}$: $l$};
		\draw[thick,->](b0.east)--(b1.west);
		\node[draw,rectangle,text width=3.5cm] (a1) at (5,0.5) {Calculate $t^l_{\text{Exit}}$, the time after which the particle leaves layer $l$.};
		\draw[thick,->] (b1.north)--(a1.south) node[pos=0.5,left]{\small $l<10$};
		\node[draw,rectangle,text width=3.8cm] (a2) at (10,0.5) {\small Calculate $\Lambda_l(t^l_{\text{exit}}v)$ using~\eqref{eq:scatterprobabilityanalytic}. Scattering inside layer $l$? I.e. is $\Lambda_{\text{total}}+\Lambda_l(t_{\text{exit}}v)>-\log(1-\xi)$? };
		\draw[thick,->] (a1.east)--(a2.west);
		\node[draw,rectangle,text width=3.8cm] (b2) at (10,-2) {\small Jump to next layer boundary.\\ Add $t^l_{\text{Exit}}\vec{v}$ to $\vec{x}$ and $\Lambda_l(t^l_{\text{exit}}v)$ to $\Lambda_{\text{total}}$.};
		\draw[thick,->] (a2.south)--(b2.north) node[pos=0.5,left] {} node[pos=0.5,right] {\small\color{myRed} No.};
		\draw[thick,->] (b2.west)--(b1.east);
		\node[draw,rectangle] (c1) at (5,-4.5) {Outside the earth.};
		\draw[thick,->] (b1.south) -- (c1.north)node[pos=0.5,left]{$l=10$};
		\node[draw,rectangle] (c3) at (10,-6.5) {Return $\vec{\Delta}=\vec{x}-\vec{x}_0$.};
		\draw[thick,->](c1.south)--(5,-6.5)--(c3.west);
		\node[draw,rectangle] (d3) at (13,-6.5) {\color{myRed}STOP};
		\draw[thick,->](c3.east)--(d3.west);
		\node[draw,rectangle,text width=3.8cm] (c2) at (10,-4.5) {\small Scattering:\\ Solve~\eqref{eq:equationnewton} for $L_s$ numerically and add $L_s \vec{e}_v$ to $\vec{x}$.};
		\draw[thick,->](a2.east)--(13,0.5)--(13,-4.5) node[pos=0.5,right]{\small}node[pos=0.5,left]{\small\color{myGreen}Yes.}--(c2.east);
		\draw[thick,->] (c2.south)--(c3.north);
	\end{tikzpicture}
	\caption{Algorithm to find the MC displacement vector $\vec{\Delta}(\vec{x}_0,\vec{v})$ of a DM particle inside the Earth.}
	\label{fig:freepathalgorithm}
\end{figure}

In conclusion, it is possible to calculate the probability of scattering after a travelled distance $L$ through a fixed layer $l$,
\begin{align}
	P_l= 1-\exp (-\Lambda_l(L))\, ,
\end{align}
where $\Lambda_l$ is an analytic function. This fact should be exploited since it will occur fairly often, that a DM particle passes an Earth layer without scattering. In that case we add up the individual terms from each layer, the particle passes through without interacting with the term of the layer where the scattering takes place,
\begin{align}
	\Lambda(L) &= \underbrace{\sum_{l}\Lambda_l (t^l_{\text{exit}}v)}_{\text{layers passed without scattering}} + \underbrace{\Lambda_{l_S}(L_s)}_{\text{layer of scattering}}\, , \quad\text{s.t. } L = \sum_{l}t^l_{\text{exit}}v + L_s\, . \label{eq:lambdasum}
\end{align}
Here, $t^l_{\text{exit}}$ it the time a particle spends inside layer $l$ before leaving. Before a scattering event the particle moves from layer-boundary to layer-boundary and at each layer change we add up the layer's contribution to $\Lambda_{\text{total}}$. Only if the particle scatters will a numerical method be used in that very layer to calculate the right term. More specifically, in a given layer $l$ there are two possibilities:
\begin{enumerate}
	\item The particle passes through the layer without interaction and we jump to the point of layer exit and add the layer's contribution to $\Lambda_{\text{total}}$,
	\begin{align}
	\Lambda_{\text{total}} \longrightarrow \Lambda_{\text{total}} + \Lambda_l(t_{\text{exit}}v).
	\end{align}
	\item The particle does scatter inside layer $l_s$, we have to solve
	\begin{align}
		\Lambda_{\text{total}} + \Lambda_{l_s}(L_s) + \log(1-\xi) = 0 \label{eq:equationnewton}
	\end{align}
	for $L_s$. Looking at~\eqref{eq:scatterprobabilityanalytic} it's obvious we have to do that numerically. Since it is a monotonous function, the Newton Method of finding roots will converge quickly. 
\end{enumerate}
Using this hybrid algorithm of analytic and numerical methods, which is also depicted in figure~\ref{fig:freepathalgorithm}, will save computation time, especially if the interaction cross-section is chosen small, such that most layers are passed without collisions. 

\section{From MC to Direct Detection Event Rates}
\label{a:DD}
In this appendix we present details of the data analysis and the statistical uncertainties. The \textsc{DaMaSCUS} simulations generate velocity data, which need to be processed to finally obtain local properties like the DM density, velocity distribution or direct detection event rates further down the line.\\[1ex]
All of the following steps are performed independently for each isodetection ring.

\subsection{Local DM Speed Distribution}
\label{ss:localdistr}
We start by describing the procedure to obtain a histogram estimate of the local velocity distribution function for each of the isodetection rings. Even though our simulations in principle provide us with the full velocity data, including the directional information, we focus on the speed distribution for now. It will be sufficient for the non-directional direct detection rate computations of this study, and it is straight-forward to extend the procedure to the full velocity distribution, which we leave for a future treatment of directional detection. Hence the goal here is to estimate the speed distribution,
\begin{align}
	g(v) \equiv \int\limits_{-1}^{1}\dd \cos\theta \int\limits_{0}^{2\pi}\dd \phi\; v^2 f(\vec{v})\, .\label{eq:gv}
\end{align}
As the simulation code tracks a DM particle on its path through the Earth's bulk mass, it records and saves its velocity $\vec{v}_i$ the moment the particle passes through one of the isodetection rings at position $\vec{r}_i$. Simulations continue until we accumulate the same statistical sample size of velocity data points $N_{\rm sample}$ for every isodetection ring. This resulting number of simulated particles is large enough that they accurately track the true underlying distribution in the underground phase space inside the Earth and we can obtain good histogram density estimates close to the isodetection ring surfaces. 

The next question is how to relate the speed data of the particles passing through the isodetection ring with the local speed distribution in the immediate neighborhood of this surface. By immediate neighborhood we refer to the surrounding volume in sufficient proximity, such that the local distribution function and density therein can be considered constant and particles will not scatter before crossing the boundary. In other word the volume size is assumed to be significantly smaller than the one defined by the local mean free path of the DM particles. We do not actually measure the particle flux in the MC simulation, i.e. the particles are not sent into the Earth continuously in time and their crossing of the rings is not time-tagged, which would be necessary to estimate the particle crossing rate. Instead we fire a number of particles and wait until each particle has run its course, waiting long enough to collect all particles passing the isodetection ring. Obviously faster particles will arrive before slow ones, but since we wait long enough we collect them all. Therefore the speed distribution constructed from these collected particles does not track the flux $\Phi$ but rather $\Phi/v$. Given that the flux is related to the distribution function (see e.g.~\cite[p.272]{reif1965}) via
\begin{align}
	\Phi(\vec{v})\dd^3v = n_{\chi}f(\vec{v})v\cos\gamma\;\dd^3v\, , \label{eq:fluxvsf}
\end{align}
 where $n_{\chi}$ is the DM number density, it is clear that the speed distribution of collected particles tracks $f(\vec{v})\cos \gamma$, where $\gamma$ is the angle between the velocity and the normal of the surface at the point of crossing. Hence in order to estimate $f(\vec{v})$ or $g(v)$ each data point needs to be weighted by the reciprocal cosine of the crossing angle $\gamma$, i.e.

\begin{align}
	w_i = \frac{1}{|\cos \gamma_i|}\, ,\quad\text{where }\cos\gamma_i\equiv\frac{\vec{r}_i\cdot\vec{v}_i}{v_i (r_{\oplus}-d_{\rm lab})}\, .
\end{align}
It is clear why we have to weigh our distribution like this in order to get the actual speed distribution. For a patch of the isodetection ring with area $dA$, a given particle sees an effective area $dA \cos\gamma$.

We estimate the distribution function underlying the data in a non-parametric way using histograms. At the beginning of the data processing we divide up the distribution's domain $(0,v_{\rm esc}+v_{\oplus})$ into $N_{\rm bins} = \lceil\frac{v_{\rm esc}+v_{\oplus}}{\Delta v}\rceil$ histogram bins $B_1=[0,\Delta v)$,~$B_2=[\Delta v,2\Delta v)$,...,~$B_{N_{\rm bins}}=[(N_{\rm bins}-1)\Delta v,N_{\rm bins}\Delta v]$. To find a suitable bin width $\Delta v$ we use Scott's normal reference rule~\cite{Scott1979},
\begin{align}
	\Delta v = \frac{3.5\sigma}{N^{1/3}_{\rm sample}}\, ,\quad \text{with }\sigma = \frac{v_0}{\sqrt{2}}\, .\label{eq:scott}
\end{align}
The height of the weighted histogram bin $i$ is given by
\begin{align}
	W_i &= \sum\limits_{j=1}^{N_{\rm sample}}w_j\, \mathbb{I}(v_j\in B_i)\, , \quad \text{where }\mathbb{I}(x\in X) = \begin{cases}
		1 \quad &\text{if }x\in X\, ,\\
		0\quad &\text{otherwise.}
	\end{cases}
	\label{eq:W}
	\intertext{Finally the weighted histogram estimation of the speed distribution $g(v)$ is simply}
	\hat{g}(v) &= \frac{1}{N}\sum\limits_{i=1}^{N_{\rm bins}}W_i \,\mathbb{I} (v\in B_i)\, ,\label{eq:gestimate}
\end{align}
where $N=\Delta v\sum\limits_{j=1}^{N_{\rm sample}}w_j$ normalizes the histogram, so that $\hat{g}(v)$ truly estimates the probability density function $g(v)$. Furthermore we determine the variance of the bin height $W_i$ based on Poisson statistics,
\begin{align}
	\sigma_{W_i}^2\simeq \frac{1}{N^2}\sum\limits_{j=1}^{N_{\rm sample}}w_j^2\, \mathbb{I}(v_j\in B_i)\, .\label{eq:histoerror2}
\end{align}
The average speed of all particles passing a certain isodetection ring is nothing but the weighted mean,
\begin{align}
	\langle v\rangle &= \frac{1}{W_{\rm total}} \sum\limits_{i=1}^{N_{\rm sample}}w_i v_i\, ,\quad \text{with }W_{\rm total}=\sum\limits_{i=1}^{N_{\rm sample}}w_i.
\end{align}
We also use the standard error (SE) approximation by Cochran (1977)~\cite{Gatz1995},
\begin{align}
	(\text{SE})^2 &\simeq \frac{N_{\rm sample}}{(N_{\rm sample}-1)W_{\rm total}^2}\times\Bigg[\sum\limits_{i=1}^{N_{\rm sample}}\left(w_iv_i-\langle w\rangle\langle v\rangle\right)^2\nonumber\\
	&\qquad\quad-2\langle v\rangle\sum\limits_{i=1}^{N_{\rm sample}}(w_i-\langle w\rangle)(w_iv_i-\langle w\rangle\langle v\rangle)+\langle v\rangle^2\sum\limits_{i=1}^{N_{\rm sample}}(w_i-\langle w\rangle)^2\Bigg]\, .
\end{align}
We performed a consistency check by running the simulations for a transparent Earth, thus without any DM-nuclei interactions. We retrieved the correct average speed for each isodetection ring, and the histogram estimates of $g(v)$ were statistically stable and successfully reproduced the Maxwell-Boltzmann distribution in perfect agreement with the Standard Halo Model, which was used to generate the initial conditions. We have also checked that our results are robust to changes in the number of simulated particles $N_{\rm sample}$.

\subsection{Local DM Density}
\label{ss:localdistr}
Next we want to extract an estimate $\hat{\rho}_{\chi}$ of the local DM number density. For this purpose we make use of two observations. The first is the fact that for free DM particles, i.e. $\sigma_{\chi n}=0$, the number density is constant throughout space and simply given by $n^{(0)}_{\chi}=\rho^{(0)}_{\chi}/m_{\chi}$ with $\rho^{(0)}_{\chi}=0.3\,\text{GeV}/\text{cm}^3$. Secondly we utilize that without the normalization the area of the histogram~\eqref{eq:gestimate} is directly proportional to the local number density in the MC simulation. 

We can therefore perform an initial run of free trajectory simulations without any scatterings and relate this to the main run including scatterings giving us the local density. For both the initial and the main simulation we sum up the weights of all particles passing through a given isodetection ring. The ratio of these two sums is proportional to the ratio of the two densities.
\begin{align}
	\frac{\hat{\rho}_{\chi}}{\rho^{(0)}_{\chi}}\sim \frac{W_{\rm total}}{W^{(0)}_{\rm total}}\, ,\quad\text{with } W_{\rm total}\equiv \sum\limits_{j=1}^{N_{\rm total}}w_j\, .
\end{align}
The superscript $(0)$ denotes the initial simulation run with no scatterings. As mentioned before, the total number of simulated particles $N_{\rm total}$ of the main run is determined by the demand of a common data sample size for all isodetection rings. Hence it will differ from the number of particles simulated in the initial run $N_{\rm total}^{(0)}$. This is taken into consideration via
\begin{align}
	\hat{\rho}_{\chi}= \frac{N^{(0)}_{\rm total}}{N_{\rm total}}\frac{W_{\rm total}}{W^{(0)}_{\rm total}}\; \rho^{(0)}_{\chi}\, .\label{eq:localrho}
\end{align}
The standard deviation is obtained by propagating the errors of $W_{\rm total}$,
\begin{align}
	\sigma_{\rho_{\chi}}^2 &= \left[\frac{\sigma_{W_{\rm total}}^2}{W_{\rm total}^2}+\frac{\sigma_{W_{\rm total}^{(0)}}^2}{\left(W_{\rm total}^{(0)}\right)^2} \right]\rho_{\chi}^2\, ,
\end{align}
where $\sigma^2_{W_{\rm total}}\equiv\sum\limits_{j=1}^{N_{\rm total}}w_j^2$. Again we point out, that these steps are done independently for each isodetection ring.

\subsection{Direct Detection Rates}
\label{ss:experiments}
Our study is relevant in case DM-nuclei cross sections are large enough that pre-detection underground scatterings can occur. These scatterings will change the DM distribution and the predictions for any direct detection experiment and detection channel. To quantify this effect we will have to give precise estimates of the local event rates based on our MC estimates described in the last section. In this work we present the case of nuclear recoil detectors as a first application.

The recoil spectrum for conventional detectors is given by
\begin{align}
	\frac{\dd R_A}{\dd E_R} &=X_A\frac{\rho_{\chi}}{m_{\chi}} \frac{ \sigma^{\rm SI}_{\chi A,\text{tot}}}{2\mu_{\chi A}^2}\eta(v_{\text{min}})\, ,\; \text{with}\label{eq:dRdER}\\
	\eta(v_{\text{min}}) &= \int\limits_{v\geq v_{\text{min}}}\dd^3v\frac{f(\vec{v})}{v} = \int\limits_{v \geq v_{\text{min}}}\dd v\frac{g(v)}{v} \, .
\end{align}
Here $X_A$ is the target mass fraction of atoms with mass number $A$. The function $\eta(v_{\rm min})$ can be calculated analytically for the Standard Halo Model, see e.g. the preprint of~\cite{Savage2007}. We already found the local DM density. The second necessary ingredient entering the event rate calculation is a histogram estimate of the $\eta$-function in~\eqref{eq:dRdER}, for which we just add up the bin areas\footnote{The bin width $\Delta v$ is the same as in~\eqref{eq:scott}.} 
\begin{align}
	H_i &= \int\limits_{v>(i-1) \Delta v}\dd v\frac{\hat{g}(v)}{v} = \sum\limits_{j=i}^{N_{\rm bins}}\Delta v \frac{\hat{g}\left((j-1/2)\Delta v\right)}{(j-1/2)\Delta v}=  \frac{1}{N}\sum\limits_{j=i}^{N_{\rm bins}}\frac{W_j}{(j-1/2)}\, ,
	\end{align}
where $W_j$ is given in~\eqref{eq:W}. This way we obtain a histogram estimate $\hat{\eta}$ for the true $\eta$-function,
\begin{align}
	\hat{\eta}(v_{\rm min}) &=  \sum\limits_{i=1}^{N_{\rm bins}}H_i\; \mathbb{I}(v_{\rm min}\in B_i)\label{eq:etahistogram}\, .
\end{align}
Together with the local DM density~\eqref{eq:localrho} we are ready to compute the MC recoil spectrum by substituting $\hat{\rho}_{\chi}$ and $\hat{\eta}$ into~\eqref{eq:dRdER}. The residual steps towards the total event rate for different experiments does not deviate from the analytic case. This study focusses on sub-GeV DM and we choose a CRESST-II type detector as an example. However we also implement a simplified computation event rate for a LUX-type detector in \textsc{DaMaSCUS} in the case one wants to study heavier DM. Both methods are described in appendix B of~\cite{Kavanagh2016}. The various integrations are done numerically using the trapezoidal rule. The uncertainty of the resulting integral is obtained by integrating the uncertainty of the integrand, which typically overestimates the statistical error.

\bibliographystyle{JHEP}
\bibliography{library}

\end{document}